\documentclass[twocolumn,showpacs,showkeys,amssymb,nobibnotes,superscriptaddress,aps,pra,bibnotes]{revtex4-1}

\usepackage{graphicx}
\usepackage{dcolumn}
\usepackage{bm}
\usepackage[T1]{fontenc}
\usepackage{mathptmx}
\usepackage{mathrsfs}
\usepackage{physics}
\usepackage{simplewick}
\usepackage{mathrsfs}
\usepackage{amsthm,amsmath,amssymb}
\usepackage[bottom]{footmisc}
\usepackage{amssymb}
\usepackage{amsmath}
\usepackage{amsfonts}
\usepackage{bm}
\usepackage{graphicx}
\usepackage{subfigure}
\usepackage[dvipsnames]{xcolor}
\usepackage{calc}
\usepackage{epsfig}
\usepackage{color}
\usepackage[utf8]{inputenc}
\usepackage[colorlinks,citecolor=blue]{hyperref}

\begin{document}
\newcommand{\be}{\begin{equation}}
\newcommand{\ee}{\end{equation}}
\newcommand{\bs}{\begin{split}}
\newcommand{\es}{\end{split}}
\newcommand{\R}[1]{\textcolor{red}{#1}}
\newcommand{\B}[1]{\textcolor{blue}{#1}}
\newcommand{\ez}[1]{\textcolor{red}{Zhou:#1}}

\title{Probing phase transition in neutron stars via the crust-core interfacial mode}

\author{Jiaxiang Zhu}
\affiliation{Center for Gravitational Experiments, Hubei Key Laboratory of Gravitation and Quantum Physics, School of Physics, Huazhong University of Science and Technology, Wuhan, 430074, P. R. China}

\author{Chuming Wang}
\affiliation{Center for Gravitational Experiments, Hubei Key Laboratory of Gravitation and Quantum Physics, School of Physics, Huazhong University of Science and Technology, Wuhan, 430074, P. R. China}

\author{Chengjun Xia}
\affiliation{Department of Physics, Yangzhou University, Yangzhou, Jiangsu, P. R. China}

\author{Enping Zhou}
\email{ezhou@hust.edu.cn}
\affiliation{Department of Astronomy, School of Physics, Huazhong University of Science and Technology, Wuhan, 430074, P. R. China}

\author{Yiqiu Ma}
\email{myqphy@hust.edu.cn}
\affiliation{Center for Gravitational Experiments, Hubei Key Laboratory of Gravitation and Quantum Physics, School of Physics, Huazhong University of Science and Technology, Wuhan, 430074, P. R. China}
\affiliation{Department of Astronomy, School of Physics, Huazhong University of Science and Technology, Wuhan, 430074, P. R. China}
\date{\today}

\begin{abstract}
Gravitational waves emitted from the binary neutron star\,(BNS) systems can carry information about the dense matter phase in these compact stars. The crust-core interfacial mode is an oscillation mode in a neutron star and it depends mostly on the equation of the state of the matter in the crust-core transition region. This mode can be resonantly excited by the tidal field of an inspiraling-in BNS system, thereby affecting the emitted gravitational waves, and hence could be used to probe the equation of state in the crust-core transition region. In this work, we investigate in detail how the first-order phase transition inside the neutron star affects the properties of the crust-core interfacial mode, using a Newtonian fluid perturbation theory on a general relativistic background solution of the stellar structure. Two possible types of phase transitions are considered: (1) the phase transitions happen in the fluid core but near the crust-core interface, which results in density discontinuities; and (2) the strong interaction phase transitions in the dense core (as in the conventional hybrid star case). These phase transitions' impacts on interfacial mode properties are discussed.  In particular, the former phase transition has a minor effect on the M-R relation and the adiabatic tidal deformability, but can significantly affect the interfacial mode frequency and thereby could be probed using gravitational waves.  For the BNS systems,  we discuss the possible observational signatures of these phase transitions in the gravitational waveforms and their detectability. Our work enriches the exploration of the physical properties of the crust-core interfacial mode and provides a promising method for probing the phase transition using the seismology of a compact star.
\end{abstract}
\maketitle

\section{Introduction}

Quantum chromodynamics is non-perturbative at low energy scales, therefore, understanding the equation of state (EoS) and the phase transition of dense matter around 1-10 times the nuclear saturation density from the first principle is an issue. Currently, models based on microscopic many-body theories and energy density functional theories are used to describe these dense matters, while there are still many uncertainties about these theories due to the complexity of the non-perturbative strong interaction. Nuclear physics experiments in the earth laboratory have a limited capability to unveil cold dense nuclear matter 
above the nuclear saturation density~\cite{Danielewicz2002_Science298-1592}.  Developing methods to probe these matter states and test the microscopic theories is important for understanding dense matter.

Compact stars, such as neutron stars \,(NS) or hybrid stars \,(HS), can carry information about these dense matters. A neutron star has a rich and interesting structure\,\cite{Caplan2017_RMP89-041002}, typically consisting of a surface ocean (which is a plasma consisting of free electrons and ions), a solid crust with dramatic density profile ($d\rho(r)/dr\sim 10^{9}\,{\rm g/cm}^4$) and a neutron-fluid core. This solid crust can be divided into two regions: the several hundred-meters thickness outer crust with density up to the neutron-drip density\,($\rho_{\rm drip}\sim 10^{11}$g/${\rm cm}^3$); the inner-crust consists of neutron-rich nuclei, free neutrons and some electrons and protons, with density ranging from the neutron-drip density to the nuclear saturation density\,($\rho_{\rm sat}\sim 10^{14}$g/${\rm cm}^3$).  In the crust-core transition region, the nuclear matter is so dense that exotic nuclear states could form, such as nuclear pasta, nuclear waffles~\cite{Baym1971_ApJ170-299, Negele1973_NPA207-298, Ravenhall1983_PRL50-2066, Hashimoto1984_PTP71-320, Williams1985_NPA435-844,Pethick1998_PLB427-7, Oyamatsu1993_NPA561-431, Maruyama2005_PRC72-015802, Togashi2017_NPA961-78, Shen2011_ApJ197-20}, etc. These exotic nuclear matter states are generated by the competition between the Coulomb force and the nuclear force.  Inside the core, for the region where the fluid density is lower than $10\rho_{\rm sat}$ but higher than the nuclear pasta density, there could also be phase transitions due to the emergence of new baryon states. In the relatively deep region of the fluid core when the density is higher than $10\rho_{\rm sat}$ and depending on the effective model of non-perturbative strong interactions, the phase transition from hadronic matter to quark matter can happen so that there will be a quark core inside the neutron star. In this case, the neutron star with a quark core is named a ``hybrid star". The onset densities for heavier baryons and quarks are still unclear with model predictions ranging from $\sim2$ to $10\,\rho_{\rm sat}$, while those new degrees of freedom are competing with each other.

These compact stars, consisting of dense matter with a density higher than the nuclear saturation density, is a natural laboratory for probing these dense matter states. Measurements of NS mass and radius play an important role in constraining the EoS of NSs~\cite{Demorest2010_Nature467-1081, Antoniadis2013_Science340-1233232, Fonseca2016_ApJ832-167, Cromartie2020_NA4-72, Fonseca2021_ApJ915-L12, LVC2018_PRL121-161101, Riley2019_ApJ887-L21, Riley2021_ApJ918-L27, Miller2019_ApJ887-L24, Miller2021_ApJ918-L28}. Recently, the discovery of gravitational waves (GWs) radiated from binary neutron star\,(BNS) systems opens a new window for probing the EoS of the dense matter: 
firstly, information such as NS tidal deformability is encoded in the inspiral GW signals~\citep{Abbott2018tidal,Abbottprx2019,Annala2017,De2018}; secondly, the electromagnetic counterparts of a BNS merger depend sensitively on the merger outcome which implies many EoS properties \citep{Ruiz2017,Rezzolla2017,Shibata2019,Margalit2017,Bauswein2017b}.

In this work, we focus on using NS/HS seismology to probe the first-order phase transitions inside NSs. A similar discussion has been proposed by Lau\,\emph{et.   al}\,\cite{Lau2021}, where they investigated the interfacial mode excited at the interface between the crystalline quark core and the neutron fluid in an HS. In their case, the typical frequency of the interfacial mode is $\sim 500$\, Hz, which is relatively high band for current ground-based gravitational wave detectors. Compared to Lau\,\emph{et. al}\,\cite{Lau2021}, our work has the following different features: (1) We focus on the interfacial mode excited at the interface between NS crust and the neutron fluid core\,\cite{McDermott1985,Tsang2012,Pan2020}, the frequency of which is typically lower frequency band, therefore, being more observationally feasible. (2) Regarding the NS, we are interested in probing the signature of the first-order phase transition of dense matter near the crust-core interface, by investigating how the crust-core interfacial mode will be affected by this phase transition. It is interesting that these first-order phase transitions could have very minor effect on the M-R relation and tidal deformability. Therefore the gravitational wave signatures associated with the interfacial mode is a promising channel to probe these phase transitions. (3) Regarding the HS with quark core, we are interested in probing the high-density  matter states by distinguishing the so-called ``twin-star'' phenomenon in the M-R relation, which is due to the introduction of hadron-quark phase transition\,\cite{Alford2013}. In this case, the two components of the twin star will have different radii and structures though their masses are the same. Therefore the interfacial mode and the corresponding GW phenomenology will also be different, which provides an observational possibility to study the phase transition of the strong-interaction matter. In addition, there are also other works\,\cite{Kruger2015} investigating the effect of phase transitions inside the crust on the oscillation mode in a hot adolescent/newborn NS in a general relativistic framework, and the effect of cooling process of a hot NS on the crust shear modes is also studied.

The structure of this paper is the following: Section\,\ref{sec:2} introduces the phase transitions in the compact star and its effect on the M-R relations; Section\,\ref{sec:3} discusses the properties of interfacial modes and how they are affected by the phase transition; Section\,\ref{sec:4} devotes to investigate the coupling of the interfacial modes to the orbital motion of binary compact stars system, and how to use the associated gravitational wave observation to probe these phase transitions; Finally, we will conclude the paper and present an outlook.

\begin{figure}[h]
\centering
\includegraphics[width=0.4\textwidth]{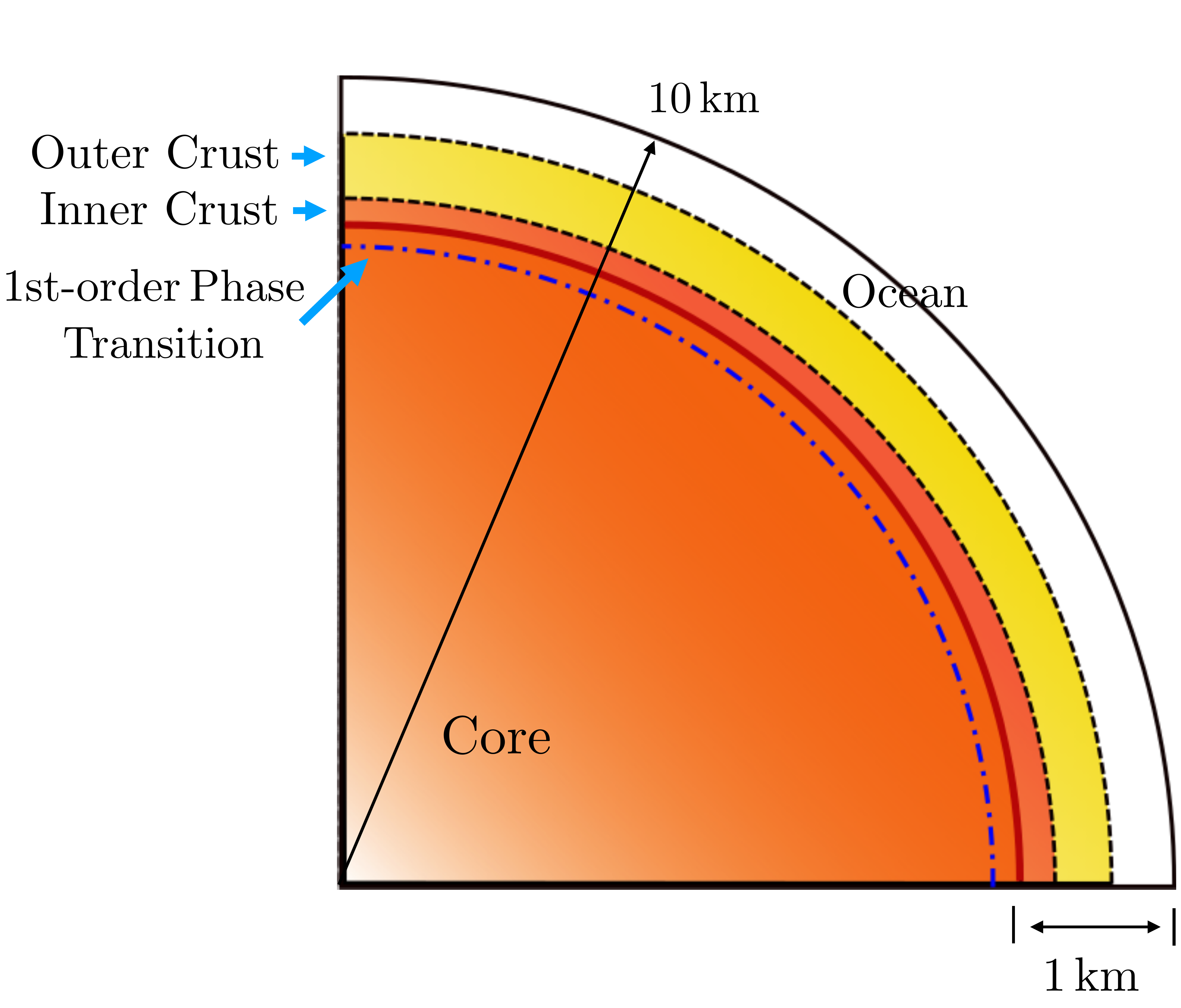}
\caption{Relevant neutron star structure for the interfacial modes are discussed in this paper. The outer crust consists of a dense Columb crystal, while the inner crust consists of dense nuclei. The nuclear pasta phase locates at the bottom of the inner crust, with only $\sim$100\, meters thickness. The density of the inner crust ranges from the neutron drip density to the nuclear saturation density. Inside the neutron star fluid core, depending on the EoS, there could be a first-order phase transition\,(denoted as blue dot-dashed lines) in the fluid core near the crust-core interface, where new baryon states could emerge. In the deeper region of the core, a first-order phase transition from the hadronic matter state to the quark matter state can happen.}
\label{fig:NS_structure}
\end{figure}

\section{Phase transitions of the high density nuclear matter}\label{sec:2}
\subsection{Hadron-Quark phase transition and the twin star scenario}\label{sec:2.1}
\begin{figure}[h]
\centering
\includegraphics[width=0.45\textwidth]{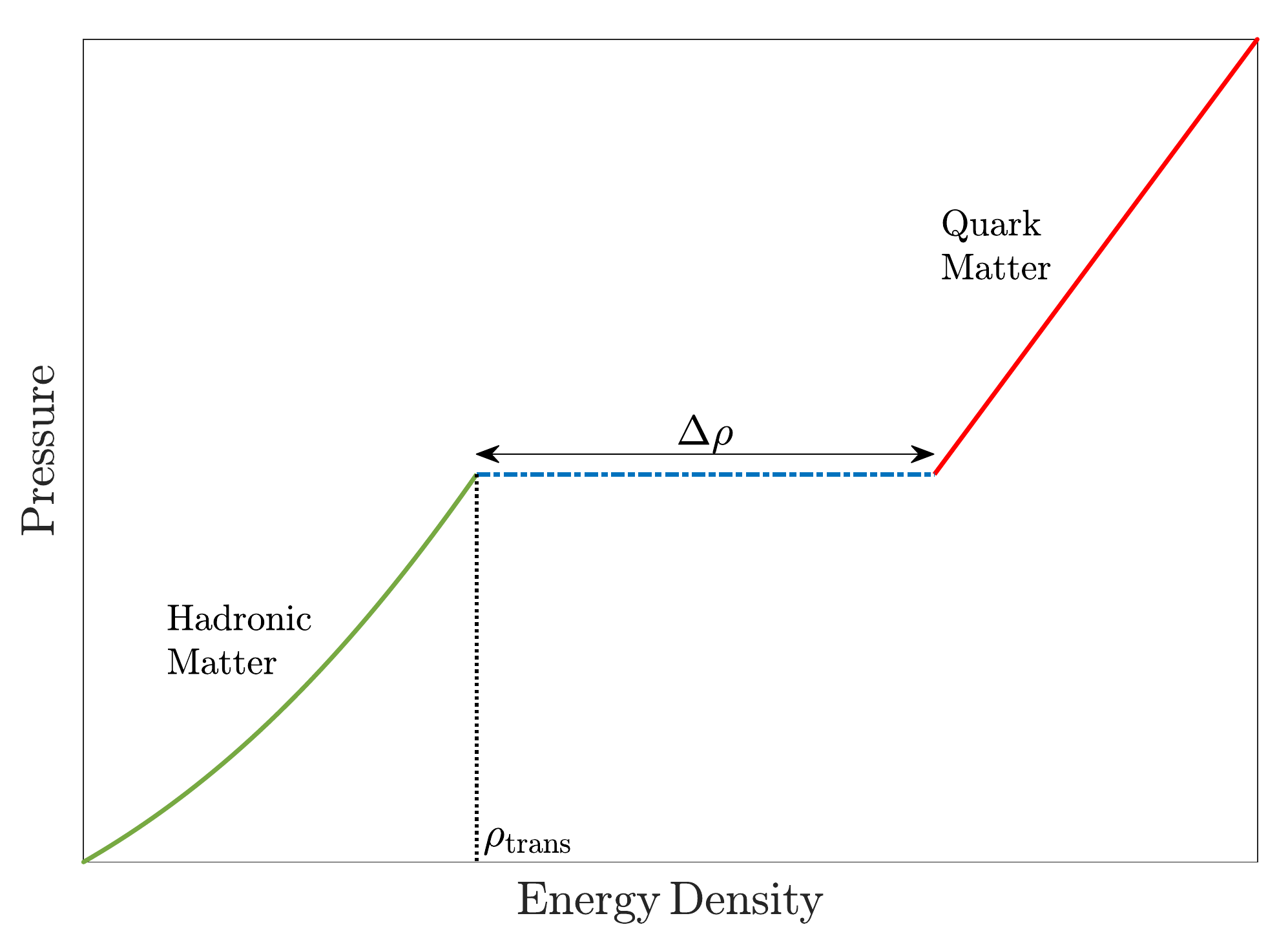}
\caption{Illustration of an equation of state model involving a hadron-quark first-order phase transition. The red line in the high energy density region represents the quark matter phase, while the green line in the low energy density region represents the hadronic matter phase. In between, there exists a first-order phase transition starting at $\rho_{\rm trans}$ with width $\Delta\rho$.}\label{fig:EoS_scheme}
\end{figure}

Phase transition in the core of a compact star can change its M-R relation. This phenomenon has been investigated in\,\cite{Alford2013} where they discussed the generic conditions for stable hybrid stars assuming that the surface tension of the phase boundary is high enough to prevent the formation of a mixed phase. Such a boundary at the interface between hadronic matter and quark matter could exist in the NS core in Fig.\,\ref{fig:NS_structure}. The general structure of the EoS we use is shown in Fig.\,\ref{fig:EoS_scheme}, where the hadronic matter phase and quark matter phase are connected by a first-order phase transition.

In the high-density region, we adopt the Constant Speed of Sound\,(CSS) model to describe the quark matter~\cite{Alford2013, Miao2020_ApJ904-103}. The CSS model has a feature that the sound speed in the quark matter is almost constant, which is independent of the baryon number density. This feature is exhibited in many microscopic quark models\,(as summarised in\,\cite{Miao2020_ApJ904-103}), e.g. the perturbative quark matter EoS\,\cite{Kurkela_2010} with a roughly constant $c_{QM}^2\approx 0.2-0.3$, the model based on the field correlator method\,\cite{Alford_2015} and some variations of the MIT bag model or NJL models\,\cite{Baym_2018,BAYM1976241}. Concretely, the CSS EoS is described by the following relation\cite{Alford2013}: 
	\be
	\begin{split}
		&\rho(p)=\rho_{\mathrm{trans}}+\Delta \varepsilon+c_{\mathrm{QM}}^{-2}(p-p_{\mathrm{trans}}), \\
		&p\left(\mu_{B}\right)=A \mu_{B}^{1+1 / c_{\mathrm{QM}}^{2}-B}, \\
		&\mu_{B}(p)=[(p+B) / A]^{c_{\mathrm{QM}}^{2} /\left(1+c_{\mathrm{QM}}^{2}\right)}, \\
		&n\left(\mu_{B}\right)=\left(1+1 / c_{\mathrm{QM}}^{2}\right) A \mu_{B}^{1 / c_{\mathrm{QM}}^{2}},
	\end{split}
 	\ee
 where $\rho, p,\mu_B,n$ are the energy density, pressure, chemical potential and the baryon number density, respectively.  The $\rho_{\rm tran}, \Delta\rho, c_{QM}$ are energy density at the phase transition point,  the energy density width of the phase transition,  and the sound speed of the quark matter, respectively. The coefficient $B=\left(\rho_{\mathrm{trans}}+\Delta \rho-c_{\mathrm{QM}}^{-2} p_{\mathrm{trans}}\right) /\left(1+c_{\mathrm{QM}}^{-2}\right)$ is determined by the quark deconfinement transition point (See the Appendix A in Ref.\cite{Alford2013} for a detailed proof), and $A$ is an undetermined constant whose value does not affect the energy density-pressure relation. For hadronic matter phase on the lower density region, we adopt the models based on the non-relativistic/relativistic mean-field theories\,\cite{Fortin2016}. By specifying that the $\mu_{B}$ remains constant after the hadron-quark first-order phase-transition, we can determine the coefficient $A$ mentioned above and thus determine the number density $n$, in this way we connect the nuclear matter and quark matter.

\begin{figure}[h]
\centering
\includegraphics[width=0.5\textwidth]{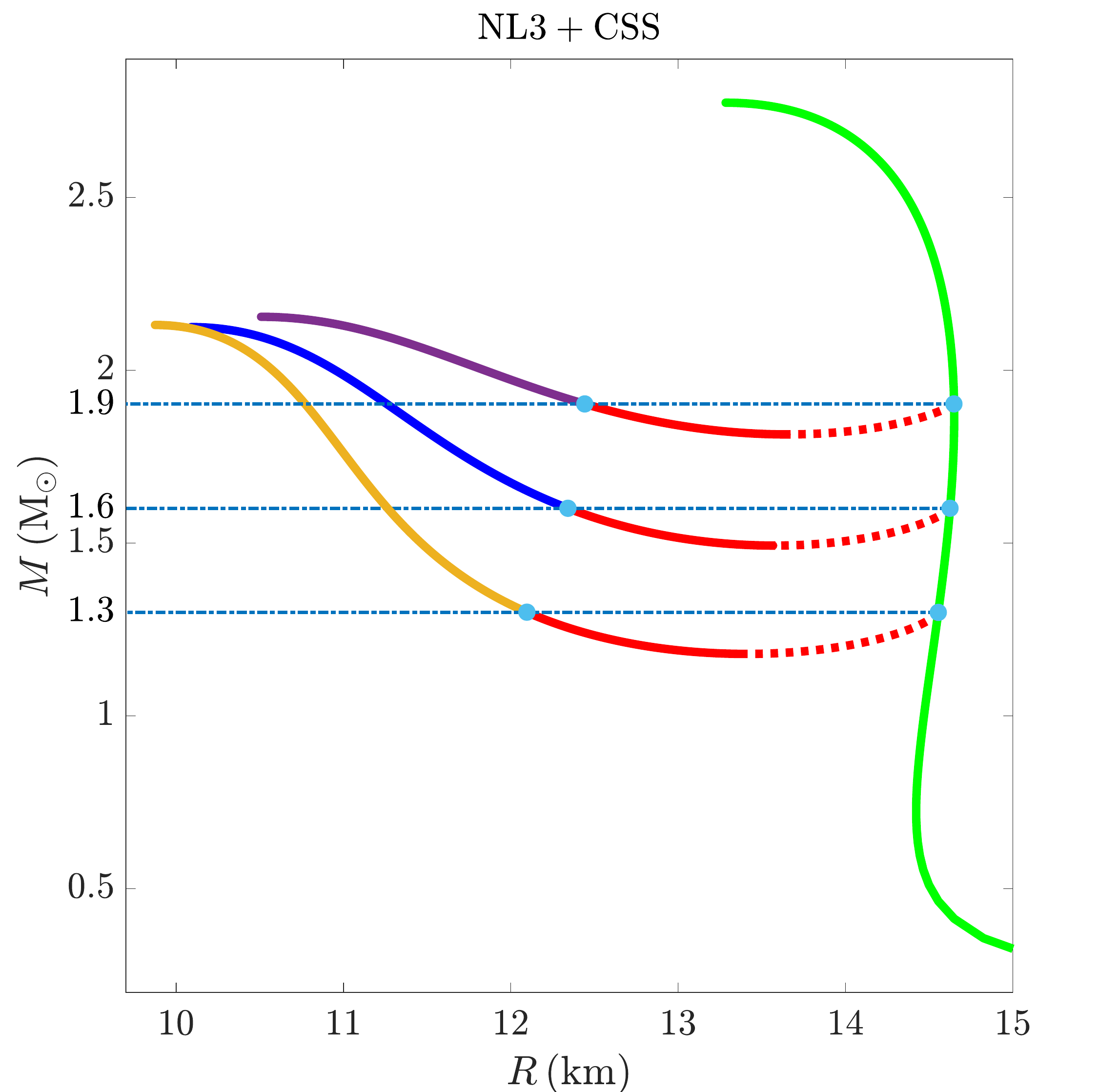}
\caption{Mass-radius relation\,(M-R) for the hybrid star with nuclear-quark transition in the core. This disconnected M-R relation exists for certain phase transition parameters with a large energy density jump $\Delta\rho/\rho_{\rm tran}$ and medium $\rho_{\rm trans}$. The twin star solution for three different masses\,$M=(1.3\,M_{\odot},1.6\,M_{\odot},1.9\,M_{\odot})$ and three different EoSs are shown. The dashed lines represent the unstable HS configuration. The EoSs of the quark core and the nuclear matter are the CSS model and NL3 model, respectively.}
\label{fig:MR_scheme}
\end{figure}

With different parameters, different M-R relations can be obtained using the above EoS and the Tolman-Oppenheimer-Volkoff\,(TOV) equations\,\cite{Alford2013}. The interesting parameter space corresponds to the so-called disconnected hybrid stars, as shown in Fig.\,\ref{fig:MR_scheme} as an example where we used the covariant density functional NL3~\cite{Lalazissis1997_PRC55-540} for the hadronic matter and the CSS model for the quark matter. For a compact star with a fixed mass, there exist two different configurations, one is a normal NS while the other is an HS containing a quark core. These twin-like compact star solutions are called the ``twin stars" scenario\,\cite{Alford2013}. The HS solution is typically more compact than the NS solution. The dashed downward cusps in Fig.\,\ref{fig:MR_scheme} correspond to the unstable solutions. This is because the constant pressure in the hadron-quark matter phase transition can not provide sufficient support against the gravity of the HS core with an increased energy density gained from the first-order phase transition. As the central density increases, the quark core can finally reach a stable state that can balance the gravitational attraction, which corresponds to the stable HS solution. The radii of twin stars with degenerate masses can be differed by several kilometres.

\subsection{Phase transition near the crust-core interface}\label{sec:2.2}
The crust-core transition region in the neutron star can be described as follows. The nuclear matter state in the neutron star crust is dominated by the Coulomb interactions among ions and the nuclear interaction between nucleons. Like ordinary materials, nuclear matter can also be classified to be ``hard matter" and ``soft matter"\cite{Caplan2017_RMP89-041002}, in terms of their elastic modulus. The NS outer crust has a solid crystalline structure, which is a hard Coulomb crystal and can also be found in the cores of white dwarfs~\cite{Brush1966_JCP45-2102, Ogata1993_ApJ417-265, Jones1996_PRL76-4572, Potekhin2000_PRE62-8554, Medin2010_PRE81-036107, Caplan2017_RMP89-041002, Caplan2018_ApJ860-148}. However, as the density increases, the competition between the nuclear attraction force and the Coulomb repulsive force starts to dominate the matter phase. Unlike the Coulomb force, nuclear force is typically short-range hence breaking the long-range order of the matter state. In this case, the usual spherical nuclear shapes become nonspherical complex shapes such as lasagna or spaghetti, and the crystal structure gradually dissolves before the nuclear matter completely becomes uniform~\cite{Baym1971_ApJ170-299, Negele1973_NPA207-298, Ravenhall1983_PRL50-2066, Hashimoto1984_PTP71-320, Williams1985_NPA435-844}.
The nuclear pasta phase is typically very thin\,($\sim 100$\,meters) and we can approximate it as a sharp interface at $r_\mu$ \emph{with discontinuous shear modulus $\mu$ but a continuous density profile}. A typical NS has a solid crust and a liquid core, therefore such an interface with discontinuous shear modulus $\mu$ always exists as long as the crust-core transition region is narrow enough. The discontinuity of the shear modulus does not affect the M-R relation. 



\begin{figure}[h]
\centering
\includegraphics[width=0.5\textwidth]{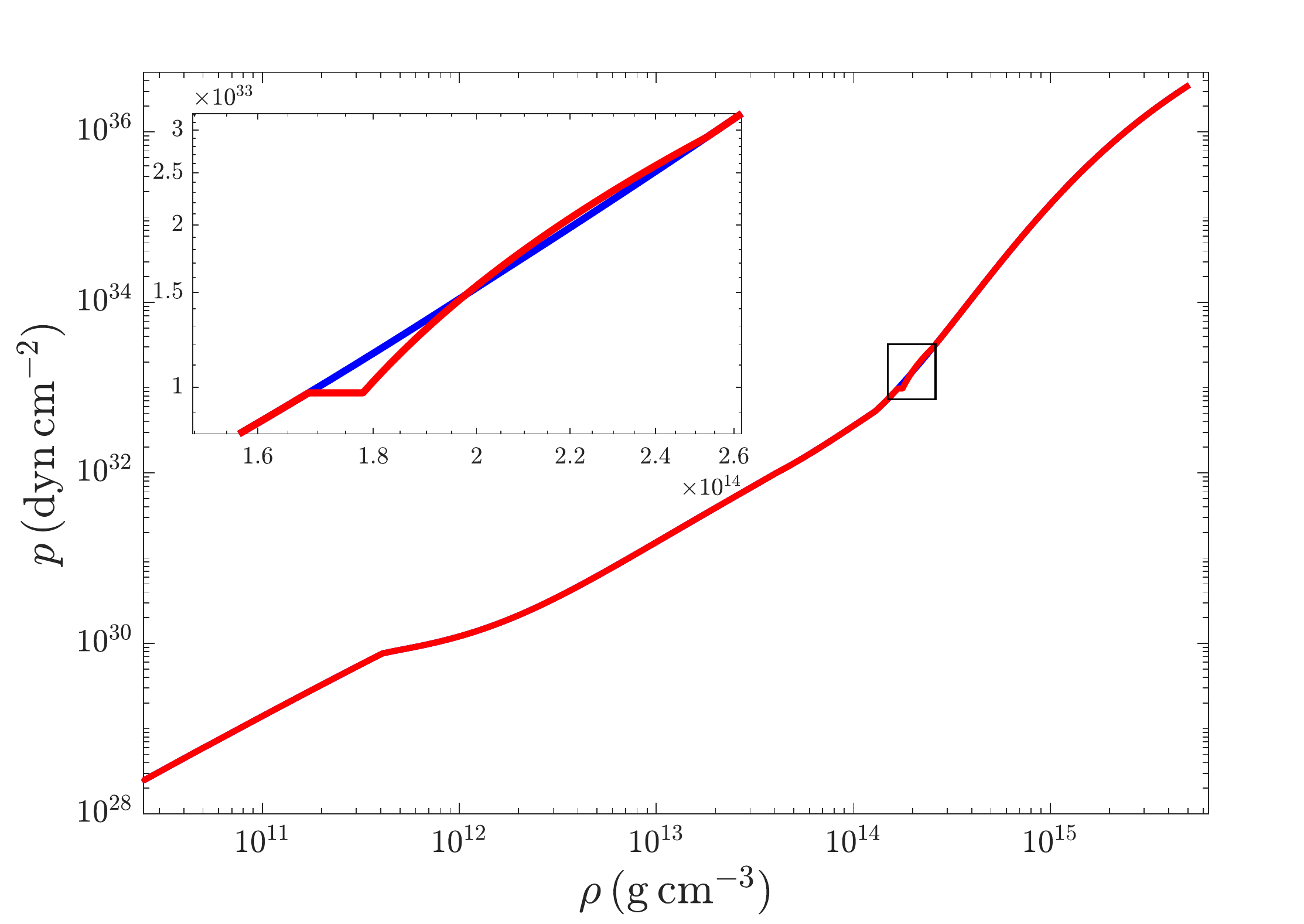}
\caption{Exemplary equation of state including the crust-core first order phase transition, where the zoomed-in figure shows the first-order phase transition region. The blue curve is the EoS without the phase transition, which is based on the SLy4 non-relativistic mean field theory model. The red curve is the EoS with phase transition constructed based on the SLy4 model, where the phase transition happens at $n_0=0.095\,{\rm fm}^{-3}$ and the relative energy density jump is $1.21\%$.}\label{fig:EoS_nuclear_scheme}
\end{figure}

In the fluid core, there could exist first-order phase transitions \emph{with discontinuous density jumps} near the crust-core transition interface. For example, we can construct such a phase transition at $r=r_\rho<r_\mu$ following the method presented in\,\cite{Fortin2016} based on thermodynamically consistent conditions, which means that the chemical potential $\mu$ on both sides of the interface are equal, while the $\partial P/\partial \mu$ and the density have discontinuity. An example of this first-order phase transition is presented in Fig.\,\ref{fig:EoS_nuclear_scheme}, which is based on the SLy4 non-relativistic mean field theory model. 

For this first-order phase transition, different viewpoints are depending on the concrete microscopic models, which are caused by the emergence of new degrees of freedom~\cite{Baym2018_RPP81-056902, Sun2019_PRD99-023004, Annala2020_NP, Dexheimer2021_PRC103-025808, Tan2022_PRD105-023018}. For example, the pion condensation can cause a transition from the normal low-density phase to the high-density phase of nuclear matter at $\rho\approx 0.9\rho_\mathrm{sat}$~\cite{Akmal1998_PRC58-1804}. 
Introducing the $\Delta$ hardon resonance in neutron stars could lead to a first-order phase transition around the crust-core transition region~\cite{Sun2019_PRD99-023004}. Bayesian inference with explicit first-order hadron-quark phase transitions from observables of canonical neutron stars shows a large likelihood (at 68\% confidence level) of phase transitions at $\rho \approx 1.6^{+1.2}_{-0.4} \rho_\mathrm{sat}$, which is also close to the crust-core transition region~\cite{Xie2021_PRC103-035802}. In general, with the stringent constraints from pulsar observations~\cite{Demorest2010_Nature467-1081, Antoniadis2013_Science340-1233232, Fonseca2016_ApJ832-167, Cromartie2020_NA4-72, Fonseca2021_ApJ915-L12, LVC2018_PRL121-161101, Riley2019_ApJ887-L21, Riley2021_ApJ918-L27, Miller2019_ApJ887-L24, Miller2021_ApJ918-L28}, the mass-radius relations at $M\lesssim 2.08 M_{\odot}$ with and without explicit first-order phase transitions should not deviate from each other significantly, indicating similar EoSs at $\rho\lesssim 6 \rho_\mathrm{sat}$~\cite{Xie2021_PRC103-035802, Jin2022_PLB829-137121, Pfaff2022_PRC105-035802}.

\begin{figure}[h]
\centering
\includegraphics[width=0.5\textwidth]{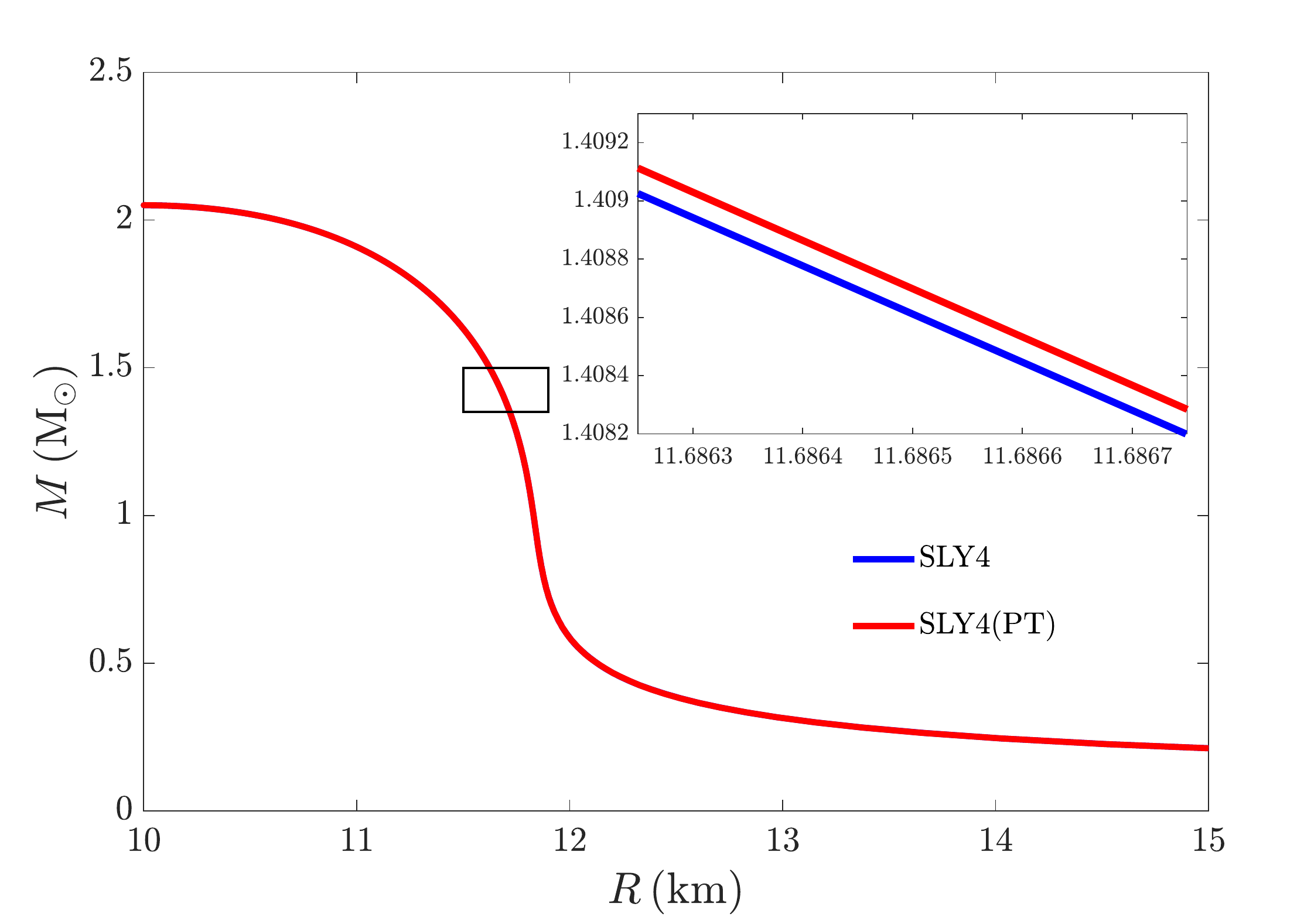}
\caption{Exemplary M-R relation corresponding to the EoS in Fig.\,\ref{fig:EoS_nuclear_scheme} based on the SLy4 model with\,(blue)/without\,(red) first-order phase transition. The introduction of the phase transition does not significantly change the M-R relation, with the small difference shown in the zoomed-in figure. This result shows that it is difficult to probe such a phase transition (if it exists) simply by measuring the M-R relation of the compact stars.}\label{fig:MR_nuclear_scheme}
\end{figure}

Furthermore, by solving the ToV equation and adopting a small density jump with the emergence of a first-order phase transition, we can also obtain the corresponding M-R relation shown in Fig.\,\ref{fig:MR_nuclear_scheme}, which means that the M-R relation is hardly affected by introducing this first-order phase transition. For the region $r>r_\mu$, that is, at the bottom of the crust, there can also be first-order phase transitions among different nuclear pasta phases. We are not interested in these phase transitions since the density jumps are often negligible\,\cite{Ravenhall1983_PRL50-2066}.

\begin{figure}[h]
\centering
\includegraphics[width=0.4\textwidth]{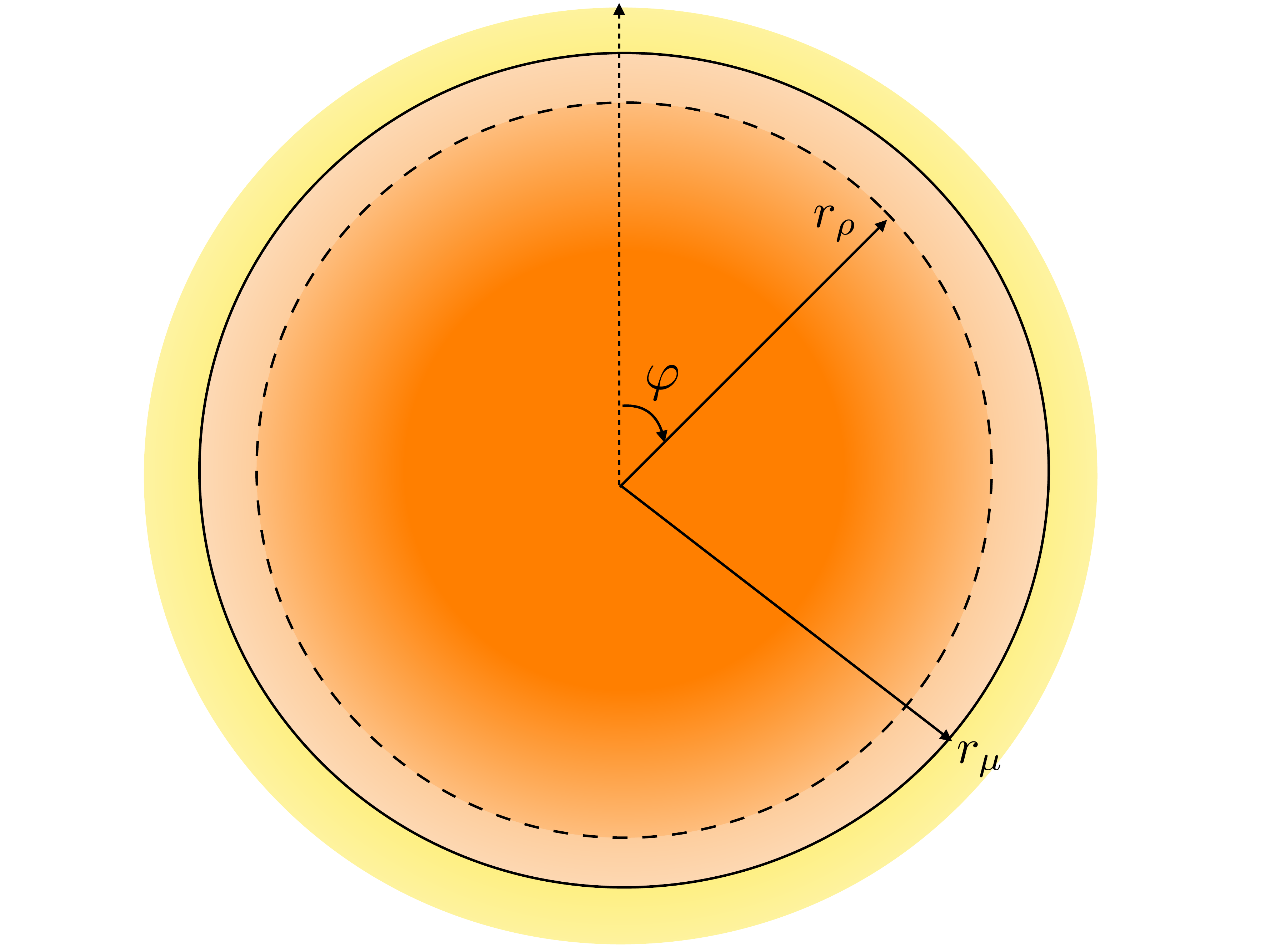}
\caption{The discontinuity of the density and the shear modulus can happen at different radii $r_\rho$ and $r_\mu$ in the compact stars. The region where $r>r_\mu$ is the crust where $\mu>0$, and the $r_\rho<r<r_\mu$ and $r<r_\mu$ regions are two fluid nuclear states with different densities. If the $r_\rho$ is in the deeper region of the fluid core and the matter with $r<r_\rho$ is quark matter state, this compact star becomes a hybrid star. The $\varphi$ denotes the azimuthal angle and the $\hat z$-axis is perpendicular to the paper.}
\label{fig:PT_position}
\end{figure}

As we shall see in the later sections, even with indistinguishable M-R relations, these phase transitions will affect the interfacial mode of the compact star, which could lead to observational signatures in the gravitational wave emitted from binary compact star systems. 

\section{Interfacial mode}\label{sec:3}
\subsection{Modes at the interfaces with shear/density discontinuity}
Under the driving of external dynamical tidal force (e.g. provided by a companion star), the internal oscillation modes of an NS can be excited  (e.g. the fundamental $f$-mode, gravity $g$-mode and pressure $p$-mode, etc.)\,\cite{McDermott1985, McDermott1988_ApJ325-725,Schumaker1983_MNRAS203-457,Hansen1980_ApJ238-740}. Among them,
one type of resonant mode takes place on the interface between different matter phases called ``interfacial modes"\,(i-mode). For example, there exists a crust-core interfacial mode\,\cite{McDermott1985,McDermott1988_ApJ325-725,Tsang2012} denoted as $i_\mu$-mode excited at the interface between the fluid core and the solid crust, where the shear modulus $\mu$ has a discontinuous transition. For the interface where a first-order transition happens with a discontinuous density jump, there will also be another interfacial mode denoted as $i_\rho-$mode\,(sometimes named as ``discontinuous-$g$-mode''). These modes can coexist in an NS shown in Fig.\,\ref{fig:PT_position}.

The coupling of the $i$-mode and the BNS orbital motion can exchange energy between the orbital motion and the internal oscillations, thereby creating observational signatures in the gravitational waves. For example, the excitation of the crust-core interfacial mode $i_\mu$ can even induce the cracking or the melting of the NS crust\,\cite{Tsang2012,Pan2020,Passamonti2012_MNRAS419-638,Passamonti2021}. Since the wavefunction and the resonant frequency of the $i$-mode are dependent on the EoS of a compact star, studying these physical phenomena related to the $i$-mode could help us understand the properties of the phases of dense matter. In the following, we will look into these interfacial modes. 

The oscillation modes of the compact star can be described by the displacement of the fluid element, represented by:
\be\label{eq:displacement}
\vec{\xi}(r,\theta,\varphi)=U(r)Y_{2,\pm2}(\theta,\varphi)\mathbf{\hat r}+rV(r)\mathbf{\nabla}Y_{2,\pm2}(\theta,\varphi),
\ee
where $U(r),V(r)$ are the displacement in the radial and tangential directions, $r$ is the NS radius. The properties of the interfacial mode were first studied by McDermott\,\emph{et. al} in the 1980s\,\cite{McDermott1985,McDermott1988_ApJ325-725}, and the relevant astrophysical phenomena were discussed by Tsang\,\emph{et al.} and Pan\,\emph{et al.}\,\cite{Tsang2012,Pan2020}\,(these modes are all $i_\mu$-modes with discontinuous shear modulus).  Following the same approach and the Cowling approximation, in this work, we calculate the interfacial mode following the Newtonian perturbation theory on a GR static background obtained by numerically solving the TOV equation, where the perturbative displacement of the fluid element satisfies:
\be\label{eq:i-mode}
(\mathcal{\hat L}+\rho\frac{\partial^2}{\partial t^2})\vec{\xi}=-\rho\nabla U_G,
\ee
in which the $\mathcal{\hat L}$ is a linear differential operator that describes the spatial dependence of the fluid displacement and the ``restoring force'' exerts on the fluid element (for details, see\,\cite{McDermott1985,McDermott1988_ApJ325-725} or the supplementary in\,\cite{Pan2020}). The term on the right-hand side of the above equation is the tidal force exerted by the companion star's tidal gravitational potential $U_G$, which will be presented in Section\,\ref{sec:4}. For solving the i-mode frequency and wavefunction, only homogeneous perturbation equations are needed: $(\mathcal{\hat L}-\rho\omega_{\alpha}^{2})\vec{\xi}_{\alpha} = 0$, where $\omega_\alpha$ is the resonant frequency of the i-mode and $\vec{\xi}_\alpha(r)=\vec{\xi}(r)e^{i\omega_\alpha t}$ is the displacement in the rotating frame of the resonant frequency. The standard shooting method is applied to obtain the eigenfrequencies of the i-mode and the mode eigenfunctions are normalised as $\int d^3x|\vec{\xi}_\alpha|^2=MR^2$. The key parameters in these perturbation equations are the $\Gamma_1\equiv d\ln{P}/d\ln{\rho}$ and the crust shear modulus $\mu$, which can be computed using the EoS table. We will introduce the two interfacial modes and their interactions in the following paragraphs

\begin{figure}[h]
\centering
\includegraphics[width=0.5\textwidth]{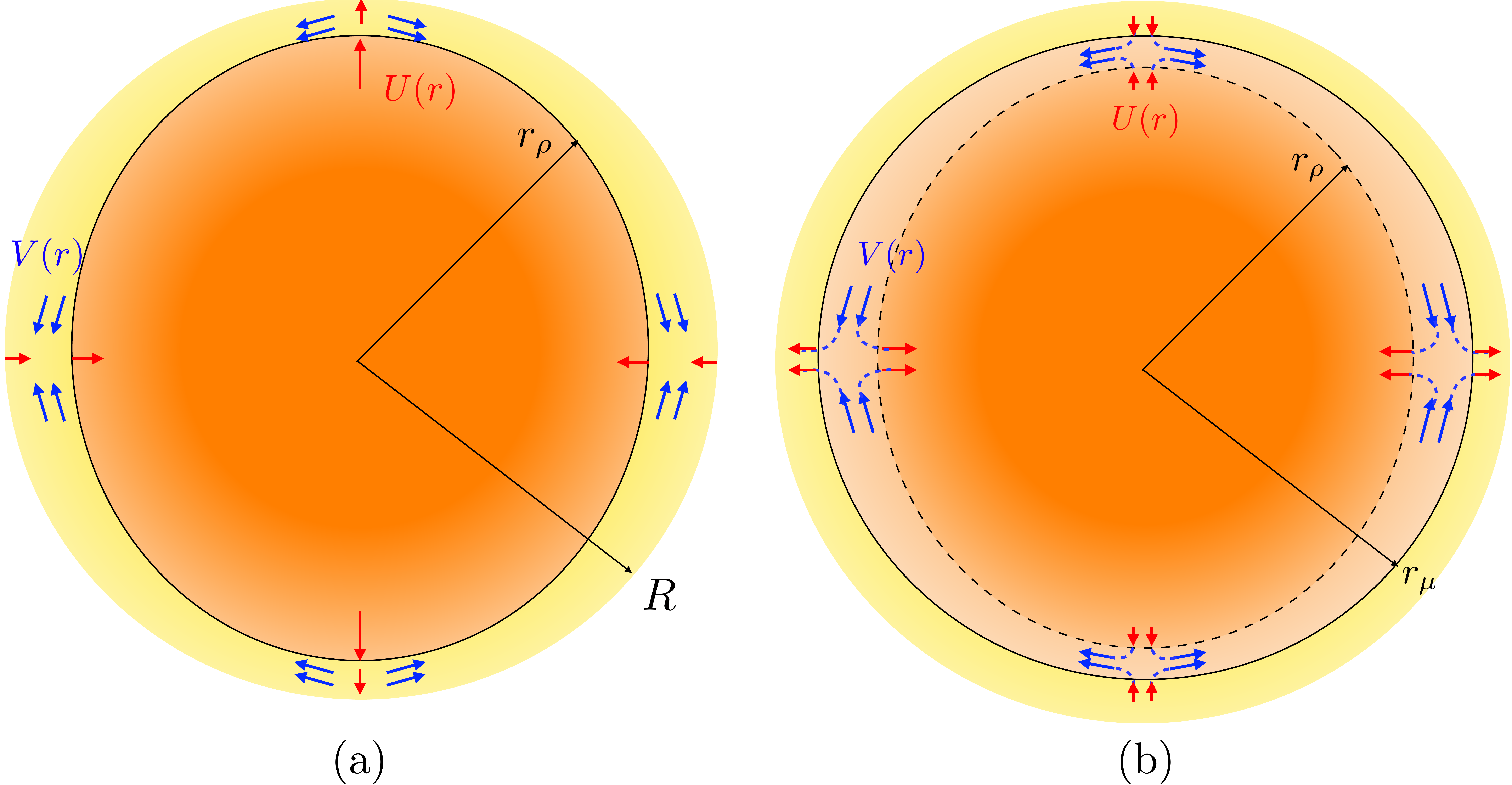}
\includegraphics[width=0.5\textwidth]{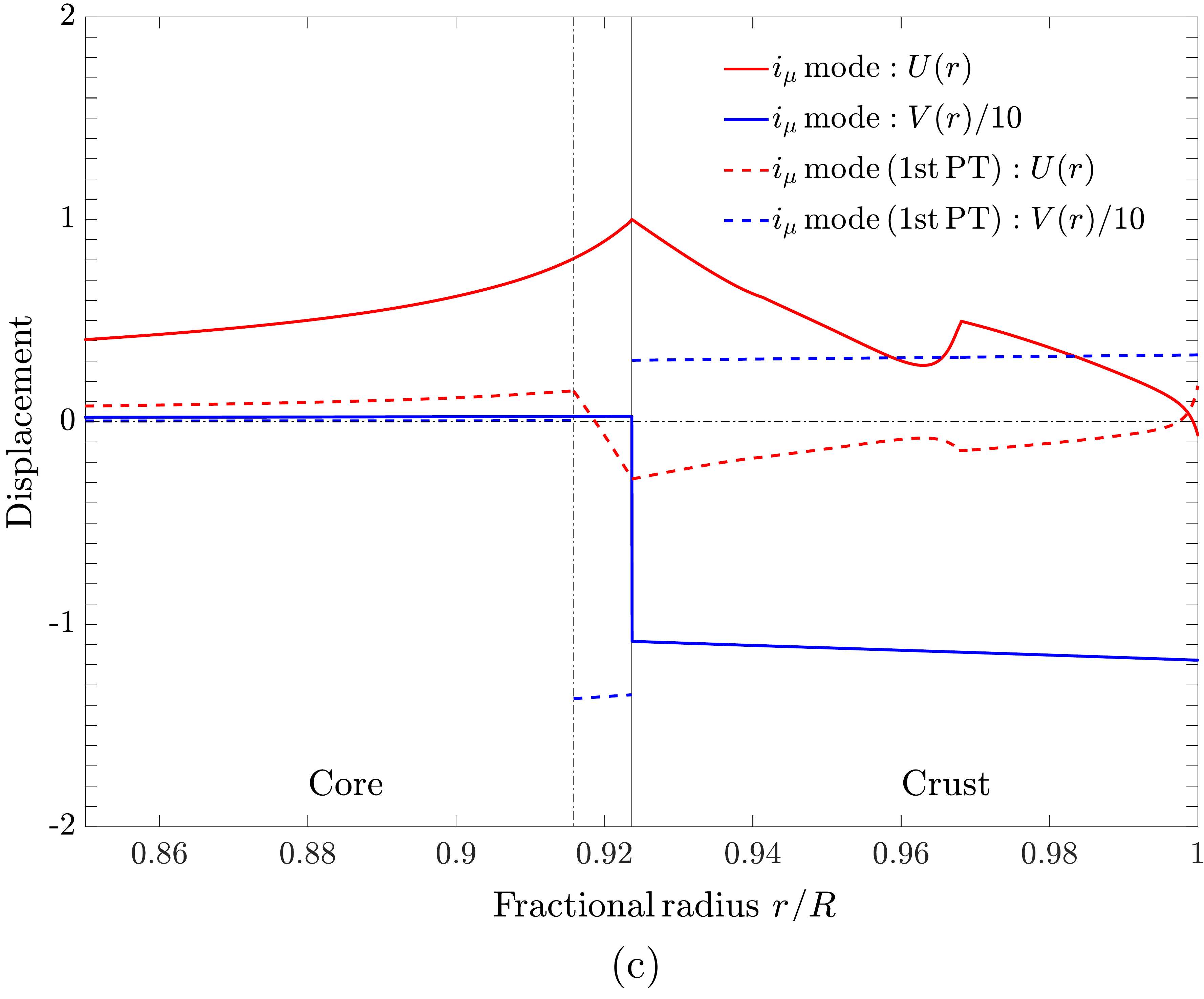}
\caption{Upper panel: Schematic diagram showing the $i_\mu$-mode of the compact star at the interface with shear modulus discontinuity. The (a/b) represents the deformation pattern without/with the first-order phase transition with density jump. The fluid within $r_\rho<r<r_\mu$ in (b) has zero shear modulus hence slipping freely. Lower panel: The interfacial mode wavefunction for the $i_\mu$-mode based on SLy4 EoS. The red/blue solid/dashed line represents the wavefunctions of $i_\mu$-mode of $U(r)$/$V(r)$ before/after we introduce the first-order phase transition with the density discontinuity. The vertical dashed and solid lines represent the density discontinuity interface and the shear discontinuity interface, respectively. Detailed physical explanations are presented in the main text.}
\label{fig:deformation_wavefunction_mu}
\end{figure}

$\bullet$ \textbf{Interfacial mode $i_{\mu}$}\,---\, Concretely, the tidal force distorts the NS core ellipsoidally in terms of $Y_{2,\pm2}(\theta,\phi)$ scalar spherical harmonics. The deformation of this core will ``squeeze" the surrounding solid crust, creating a shear deformation and compression of the nuclear matter in the crust, with the deformation pattern is shown in the Fig.\,\ref{fig:deformation_wavefunction_mu}\,(a).  We have shown the wavefunction of $i_\mu$ mode as the solid curves in the Fig.\,\ref{fig:deformation_wavefunction_mu}\,(c): the wavefunction of the radial displacement $U(r)$ has a kink at the interface; the shear displacement $V(r)$ is hardly/significantly excited in the NS core/crust, due to the discontinuity of the shear modulus crossing the crust-core interface. The directions of the shear motion of the fluid/solid element of the two sides of the interface are opposite to each other. 

With the introduction of the first-order phase transition at $r_\rho$, the interfacial mode $i_{\mu}$ is changed with the deformation pattern shown in Fig.\,\ref{fig:deformation_wavefunction_mu}\,(b). It is important to note that the fluid within $r_\rho<r<r_\mu$ has zero shear modulus and can slip freely. This means that the free-slipping fluid lyaer within $r_\rho<r<r_\mu$ will share the displacement of the crust bottom. Under the tidal deformation with $ Y_{22}(\theta,\varphi)$-pattern, at $\varphi=\pi/2$, the shear motion of free-slipping fluid element will meet and combine to push the core inwardly and the crust outwardly. Therefore the wavefunction $U(r)$ for the $i_\mu$ mode in Fig.\,\ref{fig:deformation_wavefunction_mu}\,(c) changes its sign when crossing the $r_\rho<r<r_\mu$ region.  For the shear wavefunction $V(r)$, it is also easy to see that $V(r)$ changes its sign when crossing the two interfaces. The shear displacement is relatively large within $r_\rho<r<r_\mu$ compared to the shear motion in the core and the crust, because the fluid element between the two interfaces is freely slipping during the oscillation.

The $i_{\mu}$-mode is typically soft and has a frequency around 50\,Hz, see Tab.\,\ref{tab:mode_frequency} for different EoSs. Its frequency will further decrease when interacting with the $i_{\rho}$ mode, as we shall discuss later.

\begin{figure}[h]
\centering
\includegraphics[width=0.5\textwidth]{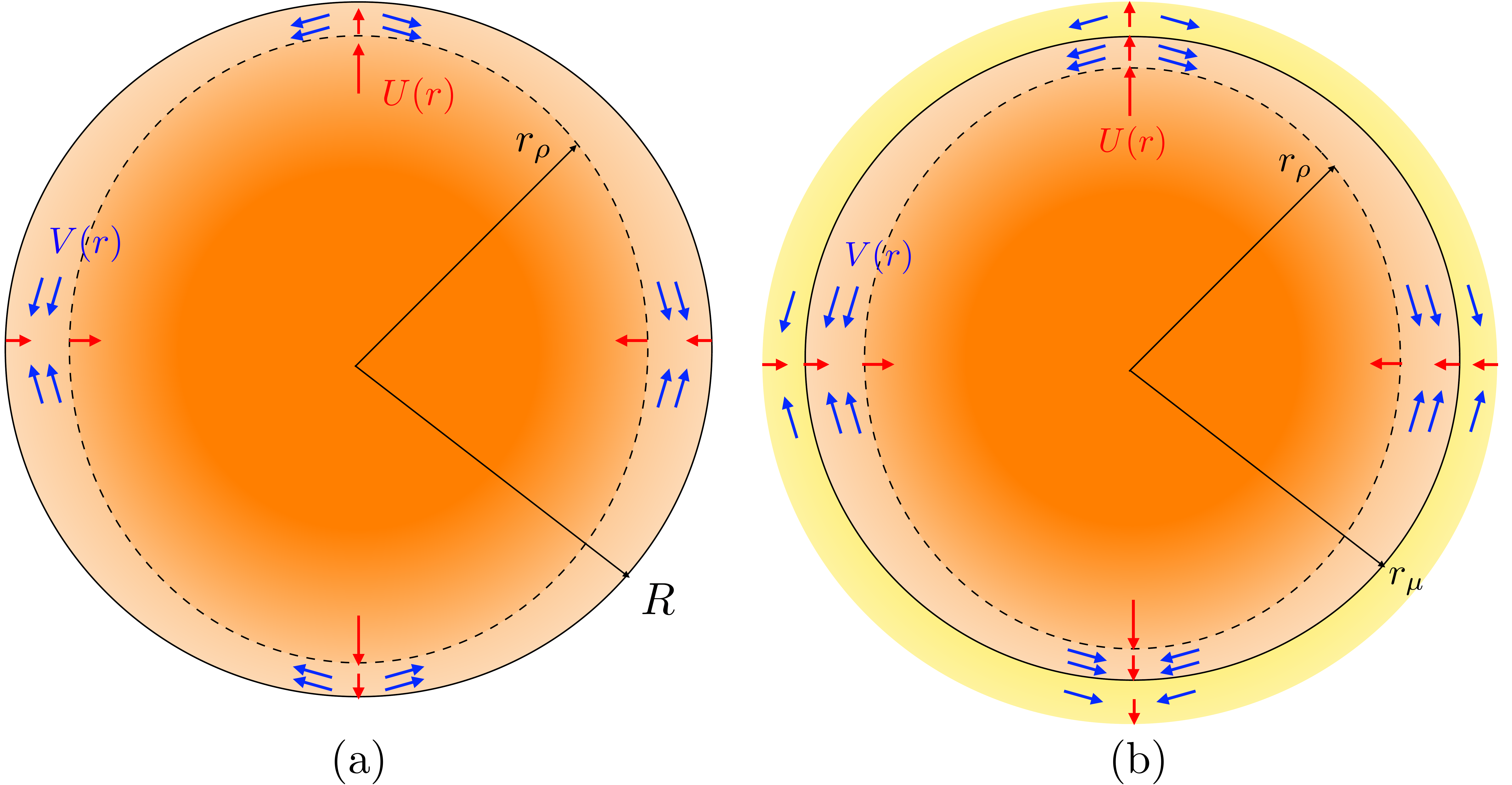}
\includegraphics[width=0.5\textwidth]{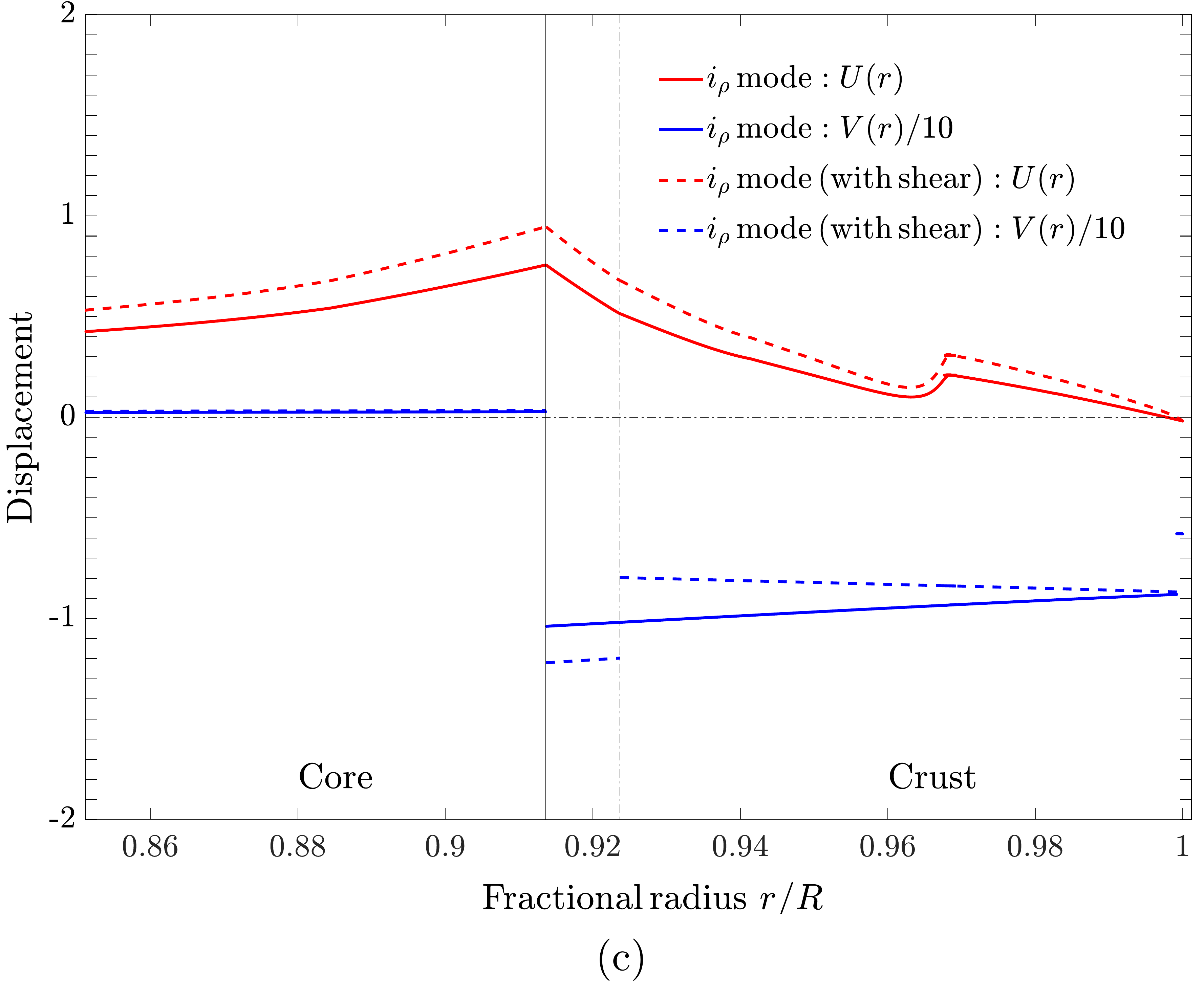}
\caption{Upper panel: Schematic diagram showing the $i_\rho$-mode of the compact star at the interface with density discontinuity. The (a/b) represents the deformation pattern without/with the shear modulus transition. The fluid within $r_\rho<r<r_\mu$ in the right panel has zero shear modulus thereby slipping freely. Lower panel: The interfacial mode wavefunction for the $i_\rho$-mode based on SLy4 EoS. The red/blue solid/dashed line represents the wavefunctions of $i_\rho$-mode of $U(r)$/$V(r)$ before/after we introduce the shear modulus transition. The vertical dashed and solid lines represent the density discontinuity interface and the shear discontinuity interface, respectively. Detailed physical explanations are presented in the main text.}
\label{fig:deformation_wavefunction_rho}
\end{figure}

$\bullet$ \textbf{Interfacial mode $i_{\rho}$}\,---\, For a fluid compact star where the shear modulus vanishes everywhere, there exists an oscillation mode $i_\rho$ excited at the density discontinuity interface, with typical deformation pattern shown in Fig.\,\ref{fig:deformation_wavefunction_rho}\,(a), and wavefunction shown as solid lines in Fig.\,\ref{fig:deformation_wavefunction_rho}\,(c). Introducing the shear-modulus discontinuity\,(or the NS crust) will modify the original ``bare" $i_\rho$-mode to that shown in Fig.\,\ref{fig:deformation_wavefunction_rho}\,(b) and the dashed lines in Fig.\,\ref{fig:deformation_wavefunction_rho}\,(c). The wavefunctions here are different from the $i_\mu$-mode case, where the radial displacement $U(r)$ does not flip the sign crossing the transition region $r_\rho<r<r_\mu$, and the shear displacement $V(r)$ only flips its sign when crossing the shear discontinuity interface. The physical picture of $i_\rho$-mode can be understood as both the crust and the $r_\rho<r<r_\mu$ fluid are passively displaced under the driving of the core deformation in $r<r_\rho$. However, in the case of the $i_{\mu}$-mode, it is the crust and the $r<r_\rho$ core passively displaced under the driving of the slipping of the fluid in $r_\rho<r<r_\mu$. Such a difference is the key to distinguishing the deformation pattern between the $i_{\mu}$-mode and $i_{\rho}$-mode.

The $i_{\rho}$ mode is relatively hard with a typical frequency larger than 100\,Hz, see Tab.\,\ref{tab:mode_frequency} for the results for different EoSs based on different nuclear physics models. The interaction with the $i_\mu$ mode will increase the frequency of the $i_{\rho}$-mode.

$\bullet$ \textbf{Mode interactions}\,---\, In the above analysis, we showed that the $i_{\mu/\rho}$-mode frequencies and wavefunctions will be altered when both density and shear discontinuity interfaces exist. This phenomenon can be understood as the mode-mode interaction, in terms of the following simple Hamiltonian:
\be
\mathcal{H}=(
\begin{array}{cc}
\xi_{\mu},\xi_\rho
\end{array})
\left(
\begin{array}{cc}
\omega_{\mu}&\chi\\
\chi&\omega_\rho
\end{array}
\right)
\left(
\begin{array}{c}
\xi_{\mu}\\
\xi_\rho
\end{array}
\right),
\ee
where the $\omega_{\mu/\rho}$ is the frequency of $i_{\mu/\rho}$ mode when only shear-modulus/density discontinuity interface exists. The $\chi$ phenomenologically describes their interactions, which depends on the radial distance between these interfaces. Typically, for a first-order phase transition that does not significantly alter the M-R relation, this distance can not be too large. Therefore the excitation of $i_{\mu/\rho}$ mode must simultaneously drive the $i_{\rho/\mu}$-mode since their wavefunctions are significantly overlapped. In this case, we will have an avoid-crossing effect for the mode-mode interaction, that is, the new eigenfrequencies become:
\be
\begin{split}
&\tilde{\omega}_{\rho}=\frac{1}{2}[\omega_\mu+\omega_\rho+\sqrt{(\omega_\mu-\omega_\rho)^2+4\chi^2}]>\omega_\rho,\\
&\tilde{\omega}_{\mu}=\frac{1}{2}[\omega_\mu+\omega_\rho-\sqrt{(\omega_\mu-\omega_\rho)^2+4\chi^2}]<\omega_\mu,
\end{split}
\ee 
which are manifested in Tab.\,\ref{tab:mode_frequency}. In obtaining Tab.\,\ref{tab:mode_frequency}, we first compute the interfacial mode $i_\mu$ or $i_\rho$ when there is only shear-modulus discontinuity or density discontinuity, respectively. After that, we find the eigenfrequencies corresponding to these two modes when both discontinuity interfaces coexists. Physically, the introduction of the free slipping layer effectively shares the excitation of the core in $i_{\mu}$-mode, causing its eigenfrequency to decrease. During the excitation of the $i_{\rho}$-mode, the introduction of shear modulus in the crust makes it harder for the core to drive the $r>r_\rho$ part of the compact star thereby increasing the mode frequency.   

\begin{figure}[h]
\centering
\includegraphics[width=0.5\textwidth]{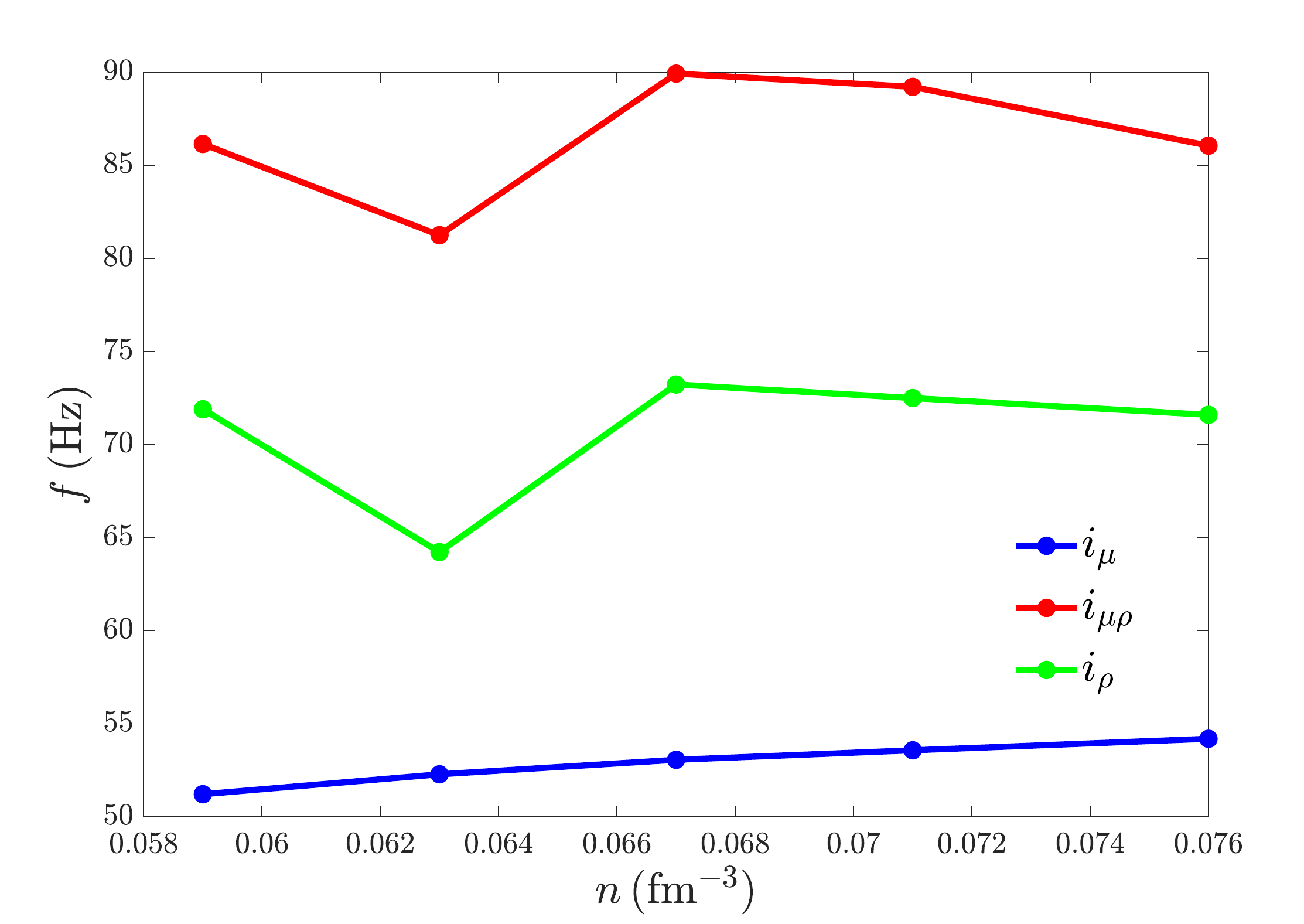}
\caption{The interfacial mode when the two interfaces coincide (using SLY4 EoS as an example): $r_\mu=r_\rho$, which we denote as the interfacial mode $i_{\mu\rho}$. In this case the fluid region $r_\rho<r<r_\mu$ vanishes and the soft mode shown in Fig.\,\ref{fig:deformation_wavefunction_mu}\,(b) disappear. The horizontal axis is the number density $n$ at which the shear modulus happens accompanied with the density jump, which serves as a free parameter due to its uncertainty. The blue/green dots are the $i_{\mu/\rho}$ frequencies when we consider the effects of the shear-modulus/density discontinuity at density $n$, respectively. The density discontinuity constructed here is rather small, ranging from 1.5\% to 2.0\%. It is shown that shifting $n$ has a minor influence on the $i_\mu$ mode frequency, while the introduction of the density discontinuity will significantly increase the mode frequency.}
\label{fig:mu_rho_degenerate}
\end{figure}

One specific case is when the two interfaces coincide, where the free-slipping fluid in $r_\rho<r<r_\mu$ region vanishes and the two modes merge into one mode denoted as $i_{\mu\rho}-\mathrm{mode}$, with frequencies shown in Fig.\,\ref{fig:mu_rho_degenerate}. Therefore in a binary compact star system, it becomes relatively more difficult for the orbit motion to drive the interface oscillation $i_{\mu\rho}$ when the shear modulus discontinuity happens at the same radius as the density discontinuity, compared to the driving of the $i_\mu$.  For the unified relativistic mean field EoSs in the CompOSE database\,\cite{2015PPN....46..633T}\cite{2017RvMP...89a5007O}\cite{https://doi.org/10.48550/arxiv.2203.03209} FSU2R, FSU2H and TM1e\,\cite{Malik_2022}, which have relatively soft inner crust structures, we find that these unified EoSs have the two interfaces coincide. In these EoSs, the crusts are connected to the core with an abrupt energy jump and a pressure plateau\,(i.e. a first-order phase transition). The connection happens at a density where the heavy nuclear clusters dissolve with a shear-modulus discontinuity. The relative energy density jumps are 16.4\%, 15.4\% and 14.2\% with the number density of the bottom of the crust being $0.083\,{\rm fm}^{-3}$, $0.087\,{\rm fm}^{-3}$ and $0.089\,{\rm fm}^{-3}$ for the FSU2R, FSU2H and TM1e, respectively. Accordingly, the $i_{\mu\rho}$-mode frequencies of $1.4M_{\odot}$ are 176.6 Hz, 177.2 Hz, and 187.4Hz. This demonstrate that these unified EoS in ComPOSE database with coincide interfaces only have one $i_{\mu\rho}$ interfacial mode with a higher frequency around $100$\,Hz.

\begin{table*}
 \begin{tabular}{|c|c|c|c|c|c|c|c|c|}
  \hline
  EoS & SKI6\,\cite{REINHARD1995467}& RS\,\cite{PhysRevC.33.335}   & SLY4\,\cite{CHABANAT1998231}     & APR4\,\cite{PhysRevC.58.1804}    & DD2\,\cite{PhysRevC.81.015803}\cite{PhysRevC.90.045803} & DDME2 \,\cite{PhysRevC.71.024312}\cite{Xia_2022}\cite{PhysRevC.105.045803}& TW\,\cite{TYPEL1999331} & DD-LZ1\,\cite{Wei_2020} \\ \hline
  $n_{\rm trans}\,({\rm fm^{-3}})$&  0.0895 &0.0672 &0.076  & 0.0672     &  0.067   &   0.074    & 0.076   &    0.07    \\ \hline
  $\Delta \epsilon/\epsilon$     &   1.47\%   &   3.41\%    &    5.19\%    &    7.36\%   &  1.52\%   &    3.30\%   &   5.06\% &   7.24\%     \\ \hline
  $f_{i_{\mu}}$\,(Hz)                   &    49.2   &  41.8     &   54.2    &    50.3     &  50.0   &   56.0    &  56.8  & 56.3       \\ \hline
  $f_{\tilde i_{\mu}}$\,(Hz)\,(1st PT)          &   23.9    &   33.4    &   36.3     &     36   &  37.2   &   37.1    &  40.3  &   36.9     \\ \hline
  $f_{i_{\rho}}$\,(Hz)                    &    81.6   &    154.3   &    152.6    &    183.7   &   91.2   &   128.7   &  166.5  &   172.5     \\ \hline
  $f_{\tilde i_{\rho}}$\,(Hz) (with shear)  &  91.9   &     156.0    &    157.7    &    186.5      &   97.6  &   135.6    &  171.4  &   177.6     \\ \hline
 \end{tabular}
 \includegraphics[width=0.55\textwidth]{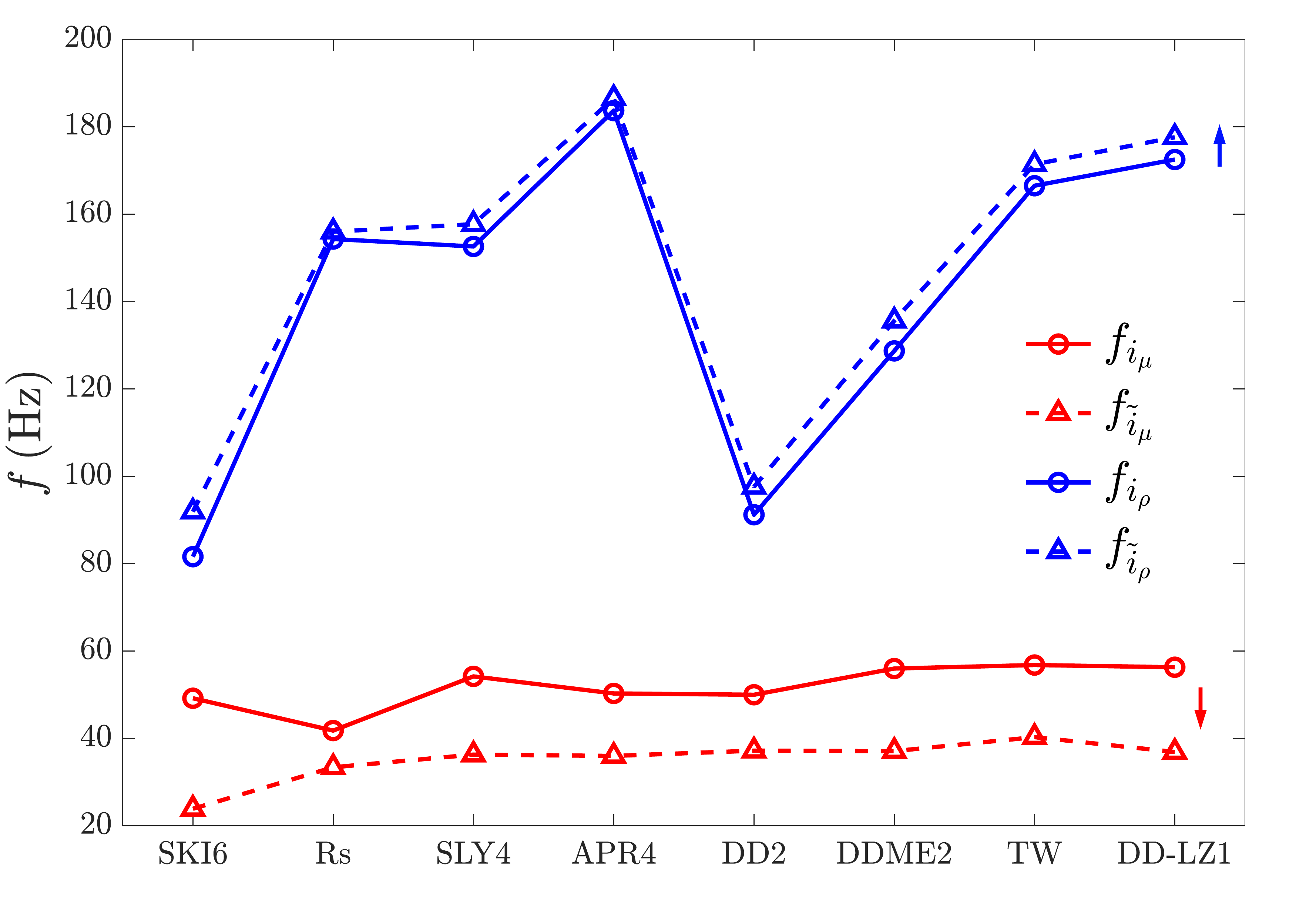}
\caption{Upper panel: The interfacial mode frequency for a 1.4 ${\rm M_{\odot}}$ neutron star computed using different EoS models partially based on CompOSE database. The density discontinuity is set to happen at $n=0.1\,{\rm fm}^{-3}$. The $n_{\rm trans}$ is the baryon number density at which the shear modulus transition happens, which is obtained from the corresponding microscopic models of the EoS\,\footnote{For the EoS data, see the ComPOSE website: \url{https://compose.obspm.fr}. In the composition table, we choose the $n_{\rm trans}$ to be the density where the proton number $Z$ and the nuclei number $A$ are zero, which is the upper limit density at which the nuclei structure disappears.}. The $\Delta \rho/\rho$ is the relative width of the first-order phase transition plateau. The $f_{i_{\mu/\rho}}$ and $f_{\tilde{i}_{\mu/\rho}}$ are the frequencies of the interfacial mode $i_{\mu/\rho}$ before and after we introduce their interactions, respectively. We can see that the interaction induces the avoiding crossing of the modes as discussed in the main text. Lower panel: Pictorial demonstration of the avoiding crossing of the modes.}\label{tab:mode_frequency}
\end{table*}

\subsection{Interfacial modes affected by the quark core}

Besides the phase transition which happens near the crust-core transition region, we have also investigated the strong interaction phase transition which leads to the existence of quark matter in the core of an HS. Lau\,\emph{et. al}\,\cite{Lau2021} analysed the interfacial mode excited at the interface between the crystalline quark matter core and a fluid hadronic envelope, which has a relatively high frequency ranging from 300\,Hz to 1500\,Hz. Intuitively, even if the quark core is not crystalline but a fluid core, there could also exist an interfacial mode (similar to the $i_\rho$-mode discussed in the previous section) at the interface of the first-order hadron-quark phase transition. Although this interface is in the relatively deeper region of the fluid core, it can still affect the properties of the interfacial mode excited at the crust-core interface with shear-modulus discontinuity. 

\begin{figure}[h]
\centering
\includegraphics[width=0.5\textwidth]{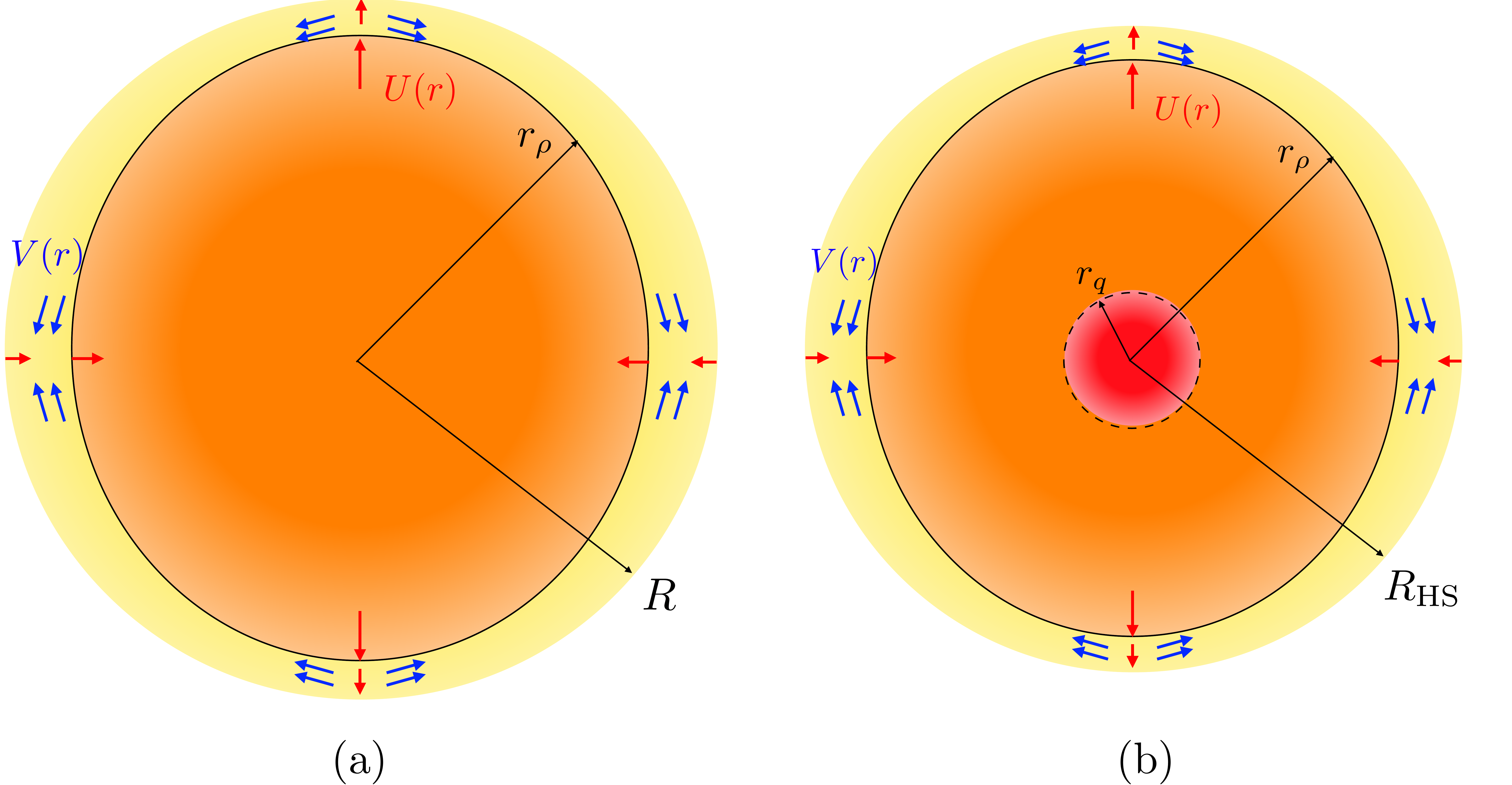}
\includegraphics[width=0.5\textwidth]{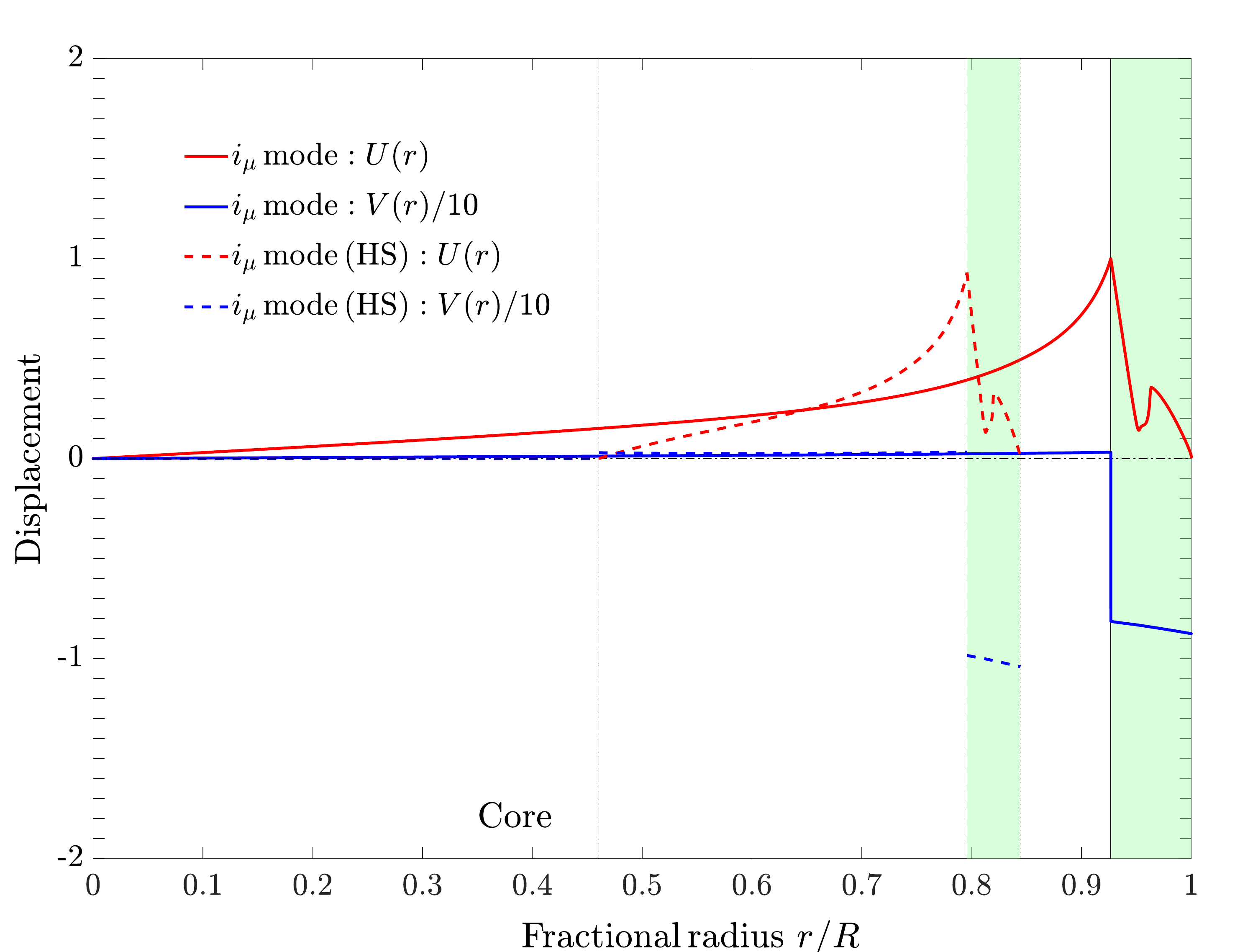}
\caption{Upper panel: Schematic diagram showing the $i_\mu$-mode of the compact star (HS and NS with the same mass $1.6\,M_{\odot}$) at the interface with the shear modulus discontinuity. The (a/b) represents the deformation pattern without/with the hadron-quark first-order transition. For the compact star with the same mass, the radius of an NS is typically larger than that of an HS and these two configurations are called ``twin-star''. Lower panel: The interfacial mode wavefunction for the $i_\mu$-mode based on SLy4 EoS. The red/blue solid/dashed line represents the wavefunctions of $i_\mu$-mode of $U(r)$/$V(r)$ before/after we introduce the hadron-quark transition. The vertical dash-dotted and dashed lines represent the hadron-quark interface and the shear discontinuity interface for the HS, respectively. The solid vertical line and the dotted vertical line are the shear discontinuity interface for the NS and the surface of HS, respectively. The $R$ is the radius of the NS and the green-shaded region is the crust of the NS/HS.}
\label{fig:deformation_twin_star}
\end{figure}

\begin{figure*}
\centering
\subfigure{\includegraphics[width=0.45\textwidth]{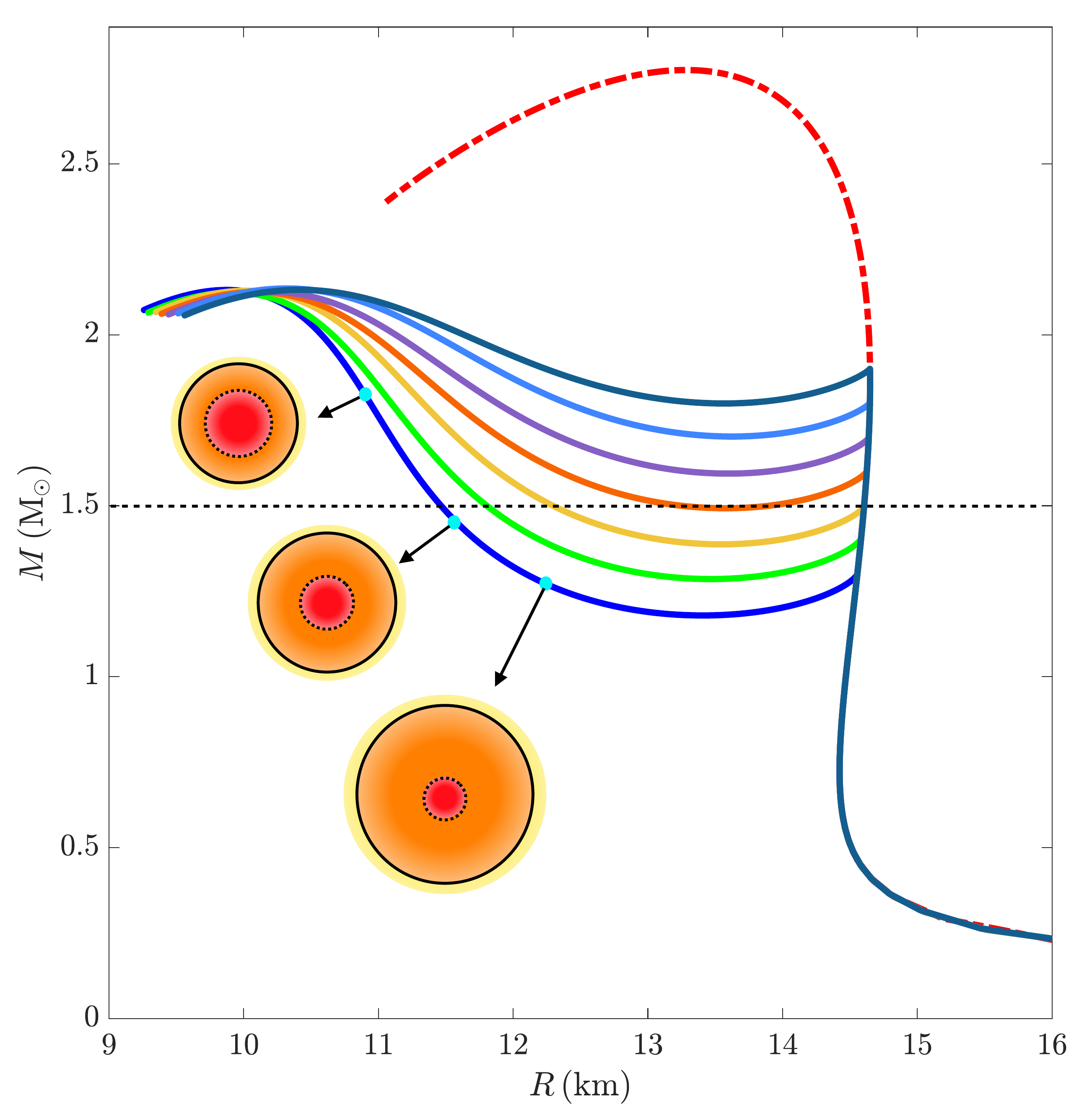}}
\subfigure{\includegraphics[width=0.45\textwidth]{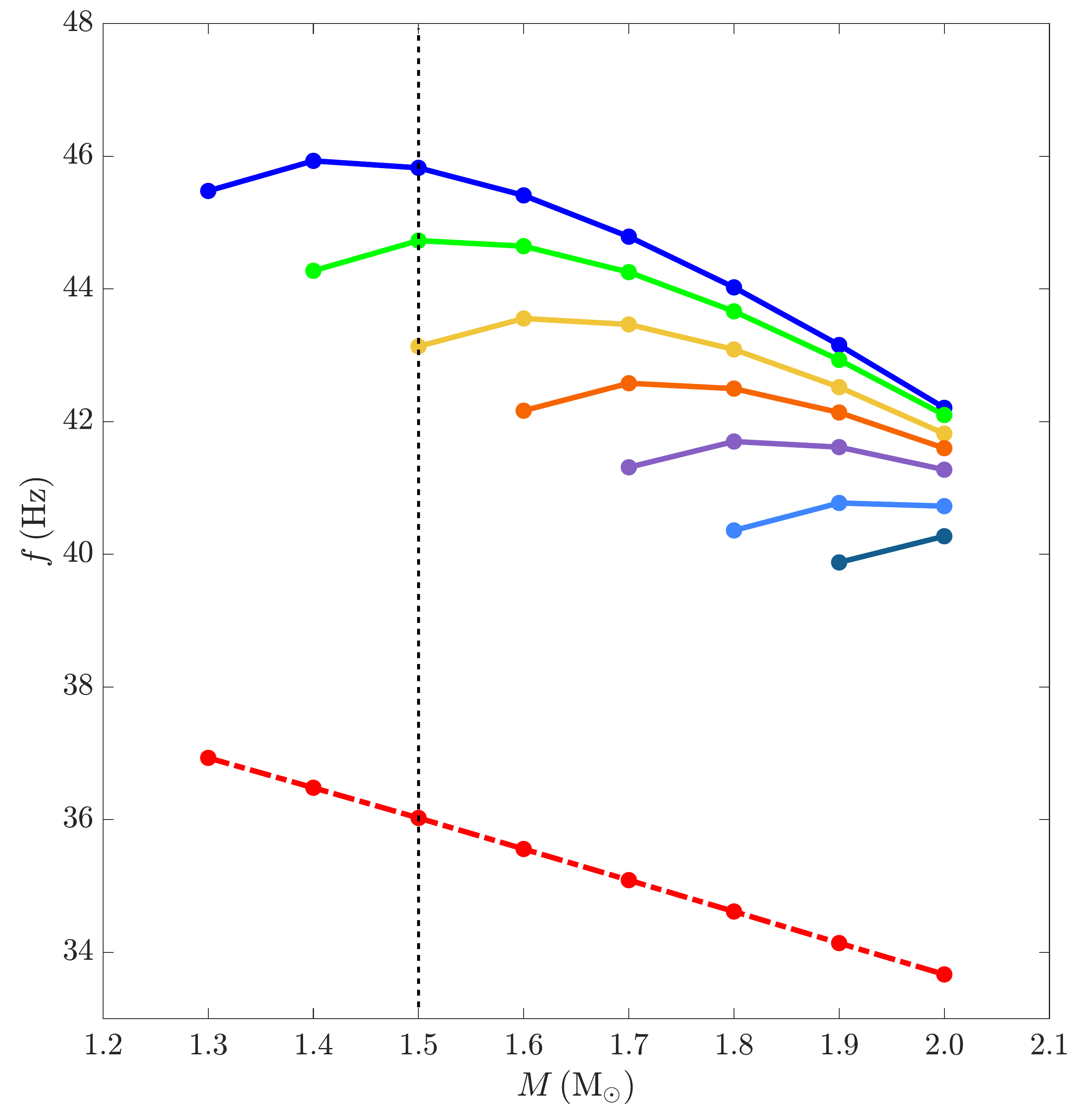}}
\caption{The interfacial mode frequency affected by the quark core. Left panel: twin star M-R relations constructed using different hadron-quark transition densities. To explore the upper limit of the twin star's effects, the speed of sound of the quark EoS is assumed to be $c$\,(which means a very stiff quark matter) and the relative energy density jump $\Delta \rho/\rho_{\rm{trans}}$ is as large as possible until reaching the observational constraints $M_{\rm{max}}\approx2.1\,M_{\odot}$. Right panel: The red dashed line represents the frequency of the $i_\mu$ mode at the NS branch (the solutions with larger radii) for different masses, and the solid line corresponds to the HS branch (the solutions with smaller radii) for different masses. The two frequencies of the two solutions in a twin-star EoS are typically differed by\,$\sim$7-10\,Hz. The EoSs of the quark core and the nuclear matter are the CSS model and NL3 model, respectively.  Detailed physical explanations are presented in the main text.}
\label{fig:twin_star_frequency}
\end{figure*}

As we mentioned in the previous section, the NS and HS with the same mass but different radii are called `'twin stars".  Interfacial modes may help us distinguish the twin branch, of which the wavefunctions are shown in Fig.\,\ref{fig:deformation_twin_star} and the frequency differences are shown in Fig.\,\ref{fig:twin_star_frequency}. We plot the frequency of the $i_\mu$-mode for seven different twin-star solutions\,(i.e. the NS and HS with the same mass) corresponding to seven different hadron-quark transition densities where NL3 hadronic EoS transits to CSS quark EoS.

For one particular hybrid star EoS, the frequency first increases and then decreases as the mass of the compact star increases. The initial frequency increase is due to the reason that, for the stable HS branch of the EoS where $dM/dR<0$, the HS radius $R$ first decreases rapidly as $M$ increases. Moreover, there also exists an interfacial mode at the hardon-quark phase transition interface discussed by Lau\,\emph{et.al}\,\cite{Lau2021}, which can also induce an avoid-crossing effect with the crust-core interfacial mode. As the mass further increases and the radius further decreases, the relative distance between the nuclear-quark phase transition interface and the crust-core interface shrinks, which creates a larger overlap between these two interfacial modes thereby a stronger avoid-crossing effect, see Fig.\,\ref{fig:twin_star_frequency}. This is why the frequency then decreases after reaching a peak value.  For one particular compact star mass (e.g. the horizontal dashed line shown in the left panel of Fig.\,\ref{fig:twin_star_frequency}), different EoSs lead to different radii of the HS, hence also different $i_\mu$-mode frequencies shown in the right panel of Fig.\,\ref{fig:twin_star_frequency}.

\section{Gravitational Wave signature}\label{sec:4}

\subsection{Interfacial mode couples to the orbital motion}
\begin{figure}[h]
\centering
\includegraphics[width=0.45\textwidth]{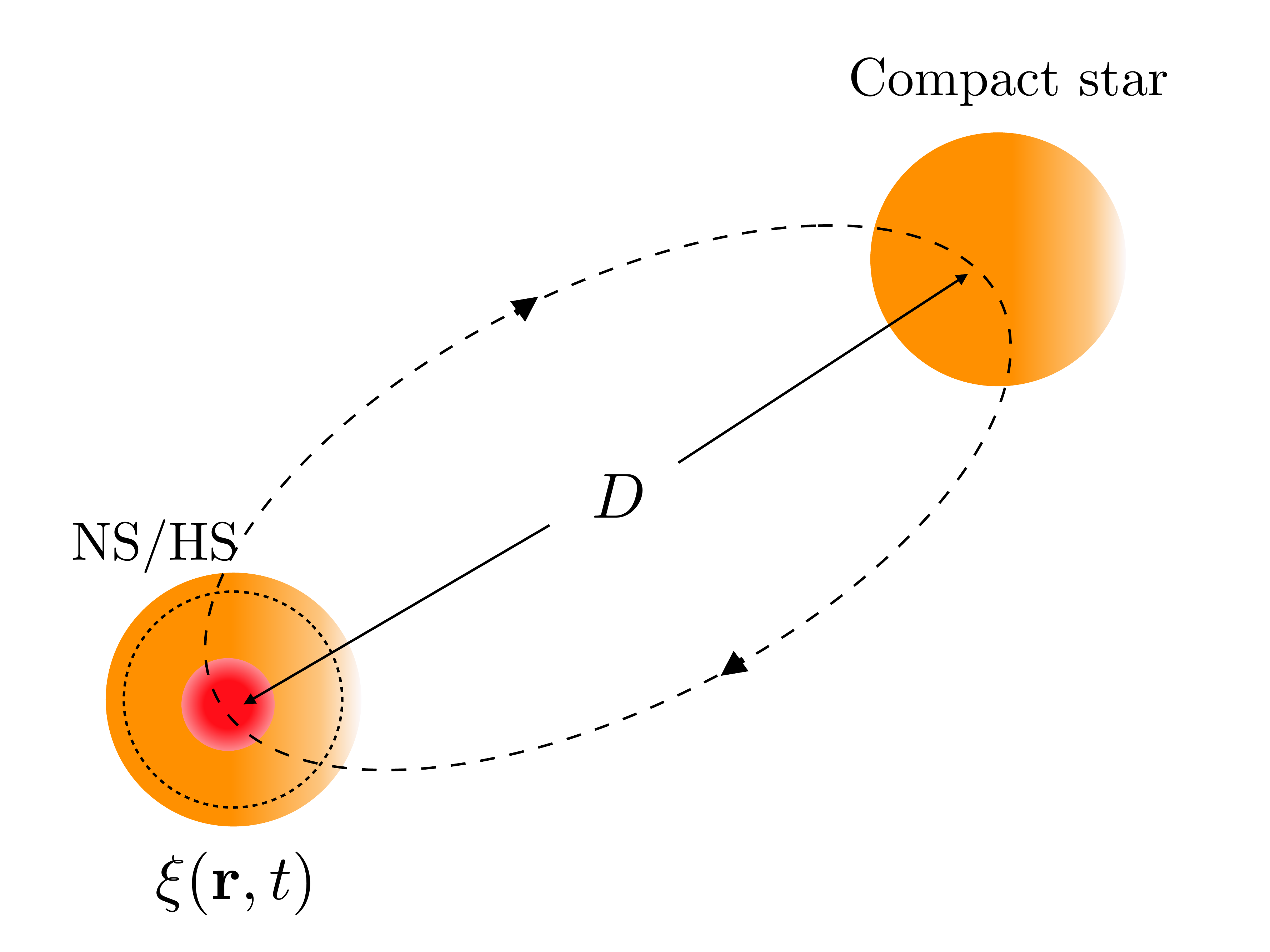}
\caption{Binary compact-stars system. The orbital motion of the binary system will exchange energy with the internal oscillation of the NS/HS through the tidal excitations. The companion compact star can be an NS, HS or even a Black hole. In this case, the interfacial mode will affect the gravitational waves emitted from the binary system.}
\label{fig:BNS}
\end{figure}

The excitation of the crust-core i-mode by the tidal force in a binary compact stars system\,(see Fig.\,\ref{fig:BNS}) will convert part of the energy of the orbital motion to the NS oscillation, thereby affecting the waveform of the gravitational waves. This phenomenon is called the dynamical tidal coupling. The internal oscillation of an NS can have many different modes, such as the pressure mode, fundamental mode, gravity mode, Rossby wave and interfacial mode. The frequencies of the pressure mode and fundamental mode are usually above the binary inspiral frequency, while the gravity mode can not efficiently couple to the orbital tidal. The Rossby wave also couples weakly since it is driven by the relativistic gravitomagnetic effect\,\cite{Flanagan2007,Sizheng2021}. The crust-core interfacial mode has a relatively lower frequency and a larger coupling to the tidal field. Therefore, the interfacial mode is of particular interest that may leave signatures in the GW emitted from the inspiral stage of a binary compact star system, and hence could be used as a probe of NS EoS and internal structures.  In this section, we discuss the impact of nuclear matter phase transitions on the GW signals emitted from the binary systems.  

Our analysis of the tidal excitation of the i-mode follows the standard method presented in\,\cite{Lai1994,Pan2020}. For the comprehensive purpose, we summarise the basic procedure as follows. The tidal excitation of the i-mode can be described by a driven harmonic equation:
\be\label{eq:tidal_driving}
\ddot{a}_m+\omega_\alpha^2 a_m+\gamma \dot{a}_m=\frac{GMW_{2m}Q_{2m}}{D^3(t)}e^{-im\Phi(t)}.
\ee
In this formula, the $a_m$ is the mode expansion coefficients defined as $\xi(\mathbf{r},t)=\sum_\alpha a_{\alpha }(t)\xi_{\alpha}(\mathbf{r})$, where $\alpha$ denotes the quantum number of the eigenmode and the $\xi_{\alpha}(\mathbf{r})$ is the neutron star eigenmode functions. The right-hand side is the tidal force derived from the tidal potential:
\be
U_G=-GM\sum_{l,m}W_{lm}\frac{r^l}{D^{l+1}}e^{-im\Phi(t)}Y_{lm}(\theta,\phi),
\ee
where the $W_{lm}$ is the spherical harmonics expansion parameter, $D$ is the orbital distance between the two NSs and the $\Phi(t)$ is the orbital phase. The spherical coordinate $(r,\theta,\phi)$ describes the position of the fluid element in the tidally deformed NS. The $Q_{lm}$ is the tidal coupling coefficient, that is, the overlap between the compact star's internal $Y_{lm}$-oscillation mode and the external tidal force, given by (for $l=2$):
\be
\begin{split}
Q_{2m}=Q^U_{2m}+Q^V_{2m}&=\frac{1}{MR^2}\int d^3x\rho\vec{\xi}^*_{n2m}\cdot\nabla(r^2Y_{2m})\\
&=\frac{2}{MR^2}\int_0^R\rho r^3dr[U(r)+3V(r)].
\end{split}
\ee 
It is important to note that the $Q_{2m}$ coefficients are different for different modes, which we listed in Tab.\,\ref{tab:tidal_coupling}. This result can be understood using the wavefunction modified by the phase transition is shown in Fig.\,\ref{fig:deformation_wavefunction_mu}\,(c) and Fig.\,\ref{fig:deformation_wavefunction_rho}\,(c). For example, the tidal coupling coefficient of the radial deformation $U(r)$ of the $i_\mu$-mode drops significantly when we introduce the first-order phase transition at $r_\rho$, because the integrand $U(r)$ flips its sign when crossing the region $r_\rho<r<r_\mu$.

\begin{table}
 \begin{tabular}{|c|c|c|c|}
 \hline
  Mode&$Q$&$Q^U$&$Q^V$\\ \hline
  $i_\mu$&0.0285&0.0425&$-0.0140$\\\hline
  $i_\rho$&$-0.0173$&0.0354&$-0.0527$\\\hline
  $i_\mu$\,(1st\,PT)&$-0.0230$&0.0001&$-0.0231$\\\hline
  $i_\rho$\,(with shear)&$0.0090$&0.0437&$-0.0347$\\\hline
 \end{tabular}
 \caption{Exemplary tidal coupling coefficients of EoS DD-LZ1 affected by first-order phase transition.}\label{tab:tidal_coupling}
 \end{table}

The $\gamma$ parameter is the dissipation rate of the i-mode, which depends on the microscopic details, and was phenomenologically discussed in\,\cite{Pan2020} and\,\cite{Lai1994}. It is worth noting that the tidal coupling to the interfacial mode could lead to some non-elastic effects. For example, Tsang\,\emph{et.al} showed that, when the shear motion of the crust reaches its elastic limit, a resonant shattering of the NS crust will happen\,\cite{Tsang2012,Passamonti2021}, which could be a possible reason for the precursor of the short Gamma-ray burst\,\cite{Tsang2012}. Moreover, Pan\,\emph{et.al} recently considered the tidal heating and the melting of the NS crust due to the excitation and dissipation of the crust-core interfacial mode, and they studied its possible signatures in the GW observations. In this case, the $\omega_\alpha$ is a time-dependent resonant frequency of the i-mode since the elastic modulus of the NS crust can decrease with an increasing temperature, and the dissipation rate $\gamma$ can also be time-dependent.  Since the crust cracking/melting process depends on the details of nuclear matter, we only qualitatively discuss these non-elastic effects in this work for an illustrative purpose later.

In computing the tidal driving on the right-hand side of Eq.\,\eqref{eq:tidal_driving}, we use the no-backaction approximation discussed and justified in\,\cite{Flanagan2007} so that the variation of the distance $D(t)$ and phase satisfies:
\be\label{eq:orbit_evolution}
\begin{split}
&\dot{D}=-\frac{64G^3}{5c^5}\frac{M_1M_2(M_1+M_2)}{D^3},\\
&\dot{\Phi}=\sqrt{\frac{G(M_1+M_2)}{D^3}},
\end{split}
\ee
which assumes that the evolution of the orbital distance is assumed to be dominated by gravitational wave radiation. When the orbital motion is on-resonance with an internal mode frequency, a tidal resonance happens. For the NS with only shear-modulus/density discontinuity in the crust-core transition region as Figs.\,\ref{fig:deformation_wavefunction_mu}\,(a) and\,\ref{fig:deformation_wavefunction_rho}\,(a), the tidal force of the inspiral stage will excite one resonance at $f_{i_{\mu/\rho}}$, of which the values are distinct from each other. For the NS with both shear-modulus and density discontinuity in the crust-core transition region as Figs.\,\ref{fig:deformation_wavefunction_mu}\,(b) and\,\ref{fig:deformation_wavefunction_rho}\,(b), there will be two resonances successively excited at different frequencies $f_{\tilde{i}_{\mu/\rho}}$  during one inspiral process. These tidal resonances will affect the gravitational wave signal and provides a way to probe the nuclear phase transition in compact stars.

\subsection{GW waveform and the detectability}
$\bullet$ \textbf{Full elastic case}--- First, we discuss the situation where there is no inelastic process in the neutron star oscillation.
The effect on the orbital cycle and hence the phase of the gravitational wave radiated by the inspiraling BNS has been studied by Lai\,\cite{Lai1994,Flanagan2007}, where the phase shift is generally given by 
\be
\Delta \phi_{j\rm GW}=2\pi \omega_{{\rm orb}}\Delta E/\dot{E}_{\rm tot},
\ee
in which $\omega_{\rm {orb}}$ is the orbital frequency on resonance with the interfacial mode, $\dot{E}_{\rm tot}$ is the energy loss rate of the entire BNS system (consists of orbital energy and the NS energy, and typically equal to the emission power of the gravitational waves), $\Delta E$ is the change of the NS energy. In case of no crust melting or shattering, the $\Delta E$ is the energy stored in the NS oscillation mode and the corresponding phase shift can be written as\,\cite{Lai1994}:
\be
\begin{split}
&\Delta \phi_{j\rm GW}=-\frac{5\pi^2}{1024}
\left(\frac{Rc^2}{GM}\right)^5\frac{2q}{1+q}|Q^j_{2,2}|^2(2\pi \tilde{f}_{j})^{-2}\\
&\approx 54\left(\frac{100\,{\rm Hz}}{f_j}\right)^{2}\left(\frac{Q^j_{2,2}}{0.03}\right)^2
\left(\frac{1.4\,M_{\odot}}{M}\right)^{4}\left(\frac{R}{10\,{\rm km}}\right)^2\frac{2q}{1+q},
\end{split}
\ee
where $q\approx 1$ is the mass ratio and $\tilde{f}_{j}=f_j\sqrt{R^3/M}$ is the normalised dimensionless i-mode frequency. In obtaining the above formula, we have considered the fact that there are two $m=\pm 2$ excited modes.  Simple estimations can show that the phase shift induced by the tidal excitation of the interfacial mode is quite large compared to the $g$ mode or $f$ mode if the excitation process is entirely elastic. If there exists an inelastic process such as crust melting, the phase shift would be much smaller as analysed in\,\cite{Pan2020} since the crust will melt before the resonant oscillation develops.

When the time-varying tidal force swept through the resonant frequencies, there will be a modification to the gravitational wave phase represented by:
\be\label{eq.phase_error}
\Delta\Phi(f)=-\sum_j\sum_{A=1,2}\delta \phi_{j\rm GW}^{(A)}\left(1-\frac{f}{f^{(A)}_j} \right)\Theta(f-f_j^{(A)}),
\ee
compared to the phase $\Phi(f)$ in the GW waveform (in the frequency domain) without this tidal resonance effect:
\be
h(f)=A(f){\rm exp}[i\Phi(f)],
\ee
where the summation over $A=1,2$ takes into account the oscillation mode of two compact stars in a binary system and the summation over $j$ accounts for the situation where there will be more than one resonant mode to be excited. For example, when the shear and density discontinuity coexists in an NS, there will be two internal modes that will be excited successively. In this case, there will be eight new parameters: $\delta\phi^{(1,2)}_{i_{\mu/\rho}}$ and $f^{(1,2)}_{i_{\mu/\rho}}$ enter into the GW waveform. For the binary system consisting of two similar compact stars, their internal modes have similar structures and properties, that is $f_i^{(1)}\approx f_i^{(2)}$, which gives the possibility to reduce the number of parameters of this model as raised in\,\cite{Pan2020}:
\be\label{eq:phase}
\Delta\Phi(f)\approx-\sum_j\delta \bar\phi_{j\rm GW}\left(1-\frac{f}{\bar f_j} \right)\Theta(f-\bar f_j),
\ee
where we define $\phi_j=\sum_A\delta \phi_j^{(A)}$ and $\bar f_j=\delta \bar\phi_j/(\sum_A\delta\phi^{(A)}_j/f^{(A)}_j)$. For the compact star with two resonant modes excited, there will be only four new parameters under this approximation. The point particle waveforms used in this work also follows\,\cite{Lau2021}, where we use IMRPhenomD template with 5\,PN and 6\,PN tidal contributions and cut-off frequency chosen as innermost stable circular orbit frequency\,(for details, see\,\cite{Lau2021}). When using these waveform templates, we recomputed the tidal deformability using the EoS modified by the phase transition is shown in Tab.\,\ref{tab:lambda}.

\begin{table}\label{tab:lambda}
 \begin{tabular}{|c|c|c|c|c|c|c|c|c|}
 \hline
EoS& SKI6& RS   & SLY4     & APR4    & DD2 & DDME2 & TW & DD-LZ1 \\ \hline
$\Lambda$ & 488&597&299&248&682&712&406&722\\\hline
$\Lambda_{\rm PT}$&496&651&303&260&714&738&424&755\\ \hline
 \end{tabular}
 \caption{The tidal deformability of different EoSs with $1.4\,M_{\odot}$. This table shows that the introduction of the first order phase transition at $r_\rho$ only induces a slight change of the tidal deformability, Tidal deformability coefficients in the template waveform is $\tilde{\Lambda}$, which is a weighted combination of the tidal deformability of the two stars in a BNS system. When $M_1=M_2$, we have $\tilde{\Lambda}=\Lambda_1=\Lambda_2$. The detection uncertainty calculated from Fisher information is $\Delta \tilde{\Lambda}\sim 100$, based on Cosmic Explorer sensitivity. This means that the tidal deformability can not be used to probe the first-order phase transition at $r_\rho$.}
\end{table}

\begin{figure}[h]
\centering
\includegraphics[width=0.48\textwidth]{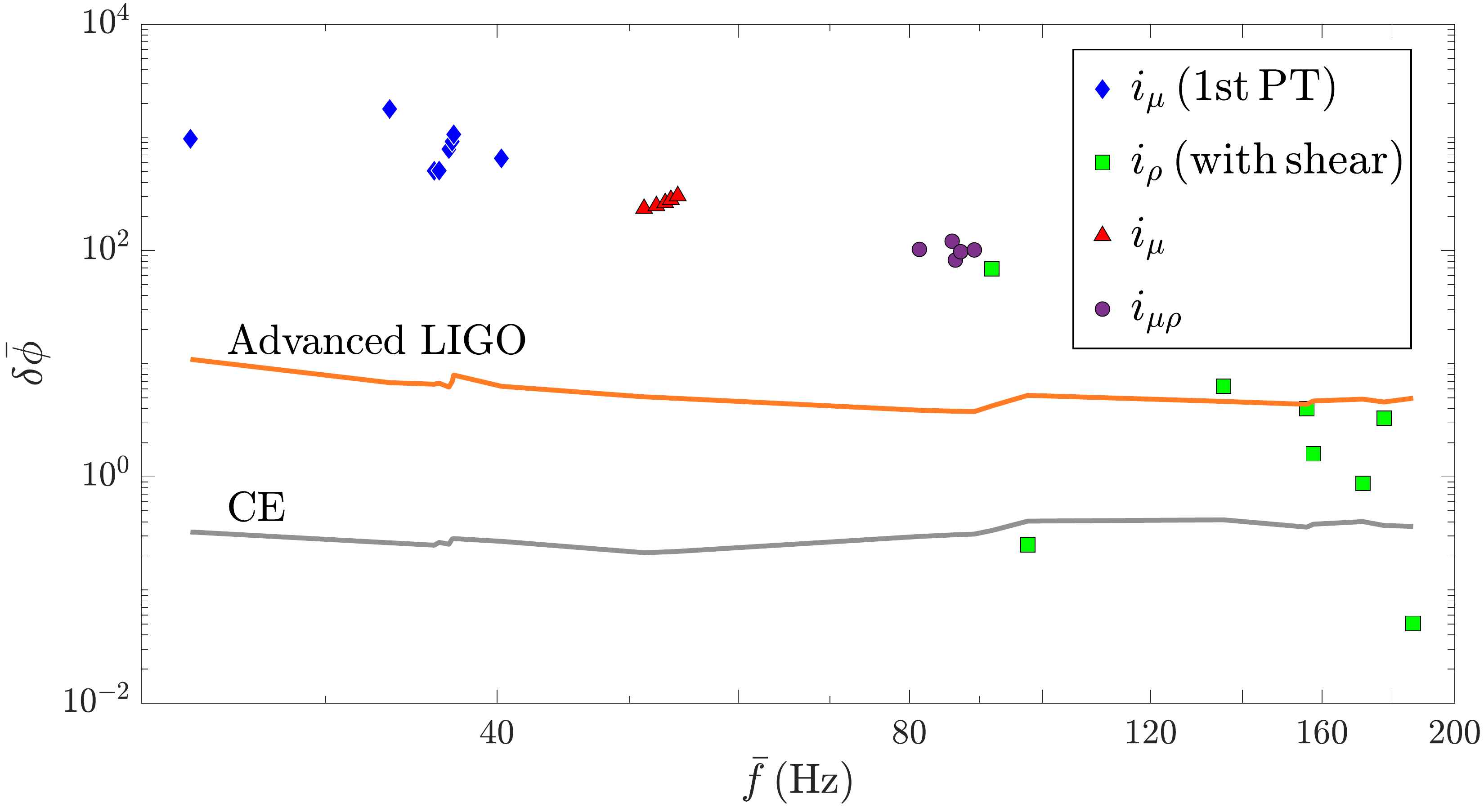}
\includegraphics[width=0.5\textwidth]{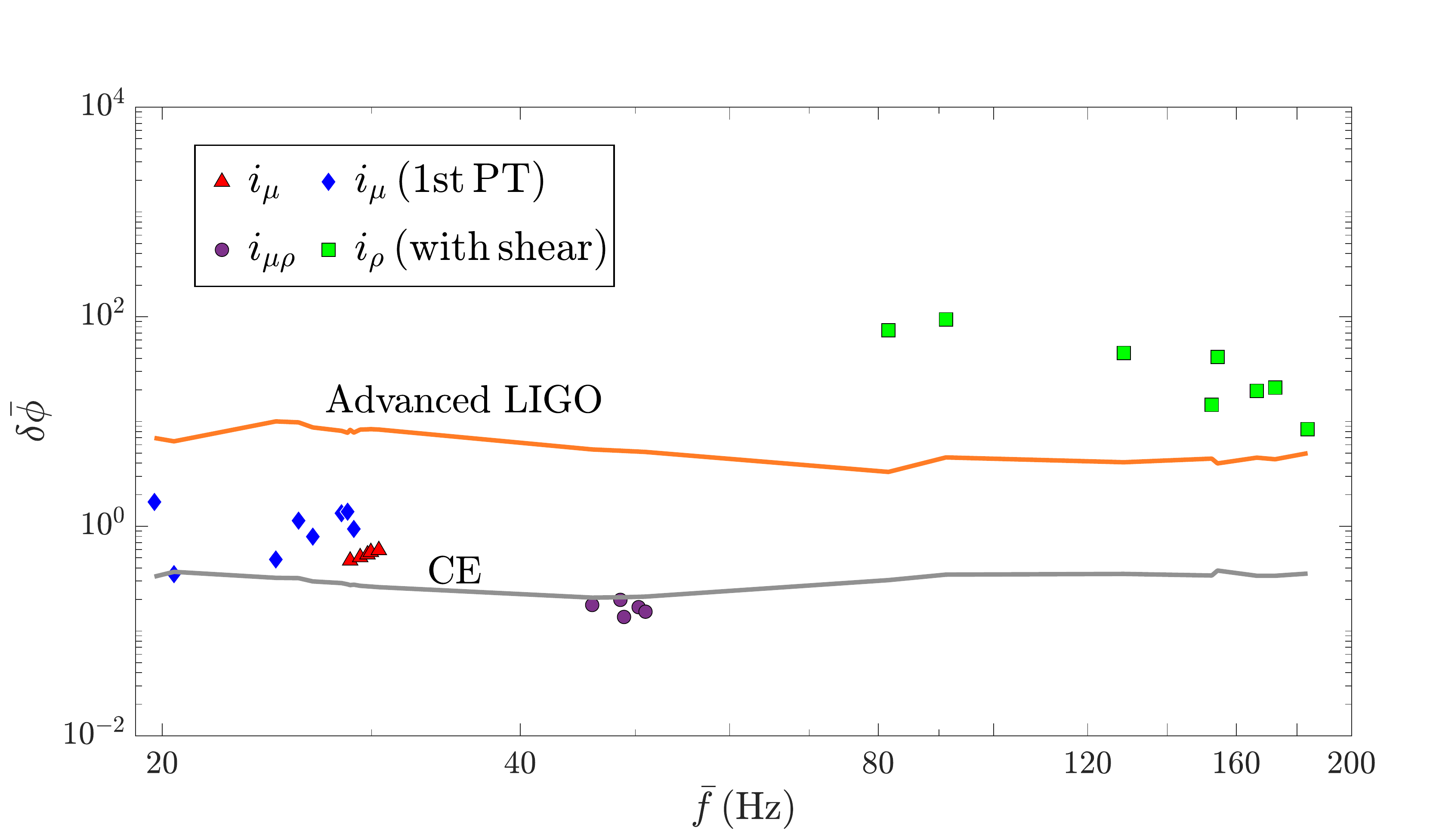}
\caption{Upper panel: Detectability of the interfacial modes for the elastic NS deformations. The vertical axis is the phase shift $\delta \phi$ due to the coupling of the orbit motion and the interfacial mode, and the horizontal axis is the redefined interfacial mode frequency for the binary system (see the text below Eq.\,\eqref{eq:phase}). The yellow and grey lines are the detectability threshold of advanced LIGO (using its designed sensitivity) and the proposed Cosmic Explorer, which are obtained using the Fisher information analysis. The blue diamond points and the green squared points describe the phase induced by the interfacial modes shown in Fig.\,\ref{fig:deformation_wavefunction_mu}\,(b) and Fig.\,\ref{fig:deformation_wavefunction_rho}\,(b), which are affected by the mode mixing. The red triangle points and circular purple points represent the interfacial modes $i_\mu$ without the effect of the density discontinuity shown in Fig.\,\ref{fig:deformation_wavefunction_mu}\,(a) and the situation when the shear and density discontinuity merges, respectively. We calculate the detectabilities for eight different EoSs in Tab.\,\ref{tab:mode_frequency}, and the mode with lower frequencies is easier to be detected. Lower panel: Detectability of the interfacial modes when the crust-melting process is introduced, the setting of the figure is the same as that of the upper panel.}.
\label{fig:detectability}
\end{figure}

We model the detectability using the Fisher information matrix as in\,\cite{Cutler1994,Yu2017,Lau2021}, supposing that the gravitational wave signal is strong enough to have a high signal-to-noise ratio, and the detector noise is Gaussian. The effective Fisher matrix incorporating $a\,priori$ information following \cite{Cutler1994} with template parameters $\{\theta_m\}$ defined by:
\be
\tilde{\Gamma}_{mn}=\left(\frac{\partial h}{\partial \theta_m}\vline\frac{\partial h}{\partial \theta_n}\right) + \frac{\delta_{mn}}{\sigma^2_{m\,{\rm prior}}},
\ee
where the inner product $(a|b)$ is defined as
\be
(a|b)=2\int^\infty_0\frac{a^*(f)b(f)+a(f)b^*(f)}{S_{hh}(f)},
\ee
with $S_{hh}(f)$ is the strain noise spectral density of the GW detector.
The root-means-square\,(rms) for estimating the parameter $\theta_m$ now can be written as:
\be
\Delta \theta_m=\sqrt{({\bm \tilde{\Gamma}}^{-1})_{mm}},
\ee
where $\bm \tilde{\Gamma}^{-1}$ is the inverse of the effective Fisher information matrix. The phase shift is detectable if it is larger than the $\Delta\delta\bar\phi_{j\rm GW}$. In Fig.\,\ref{fig:detectability}, we plot the detectability of the interfacial modes computed using different EoSs, based on the designed sensitivity of advanced LIGO\,\cite{Aasi_2015} and the 3rd generation detector Cosmic Explorer\,\cite{Dwyer2015}. For the $i_\mu$-mode at the shear discontinuity interface with lower frequency, there is a significant detectability of these modes from the GW observations if the excitation process is entirely elastic; while the interfacial mode at the density discontinuity interface is relatively more difficult to be detected due to its high frequency. The introduction of the first order transition reduced the frequency of the $i_\mu$-mode while increases its detectability.

$\bullet$ \textbf{Influence of the inelasticity to the detectability}---The results presented in Fig.\,\ref{fig:detectability} show that for the elastic deformation, there is a significantly high detectability. However, as we mentioned before, Tsang and Pan\,\cite{Tsang2012,Pan2020} investigated the possible inelastic process that happens in the deformation of an NS crust. In particular, Pan\,\cite{Tsang2012,Pan2020} considering the melting of the crust and the basic physical process can be described as follows: the tidal interaction drives the i-mode so that the shear motion may exceed the crust yield limit and the plastic deformation starts, accompanied with the mode energy converting into the thermal energy. Gradually accumulating heating energy will start to melt the crust, and it takes $\sim 20$ orbit periods to completely melt down the crust which costs about $10^{47}$\,ergs energy. Pan\,\cite{Tsang2012,Pan2020} gave the formula for estimating the phase correction induced by the crust melting as:
\be\label{phimelt}
\begin{split}
&\delta\phi_{\rm melt}=\frac{2\omega_{\rm orb}E_{\rm melt}}{P_{\rm GW}}\approx\\
&\frac{0.18}{q^2}\left(\frac{1+q}{2}\right)^{2/3}\frac{E_{\rm melt}}{10^{47}{\rm erg}}\left(\frac{M}{1.4\,M_{\odot}}\right)^{-10/3}\left(\frac{f_{\rm GW,melt}}{100\,{\rm Hz}}\right)^{-7/3},
\end{split}
\ee
which can be used to estimate the effective melting to the detectability. Using the approach as in Pan\,\cite{Pan2020}, we plot the effect of melting on the detectability of interfacial modes for different EoS models in the lower panel of Fig.\,\ref{fig:detectability}, where we considered the temperature-dependent shear modulus and the detailed crust heating model\,(summarised in the Appendix for illustrative purpose). The results in Fig.\,\ref{fig:detectability} can be understood as follows: (1) The effects on the GW phase due to the coupling of the orbital motion with the $i_\mu$ modes become more difficult to observe since the melting destroys the crust, and the melting energy $E_{\rm melt}$ is only a small fraction compared to the resonant energy when the $i_\mu$ mode is fully swept through during resonance. (2) The resonant frequency for the driving of the crust-core interfacial modes $i_\mu$ systematically move leftward because the heating soften the crust matter. Moreover, since the exceeding of elastic deformation happens before the original resonant point, the accumulation of heat and the meltdown of the crust effectively lower the resonant frequencies, which is beneficial to the observability of the $i_{\mu\rho}$ mode (see Fig.\,\ref{fig:detectability}). (3) The melting effect will convert the crust from a Columb crystal to a hot fluid, so that the discontinuity of the shear modulus in the cold NS no longer exists. Despite the difference in chemical composition, the matter phase in the $r_\rho<r<r_\mu$ region and that within $r_\mu<r<R$ have no difference. Finally, the coupling between the $i_\mu$ and $i_\rho$ also diminishes, which will create a decrease of the $i_\rho$ frequency and an increase of the tidal coupling coefficient $Q$ (see Tab.\,\ref{tab:tidal_coupling}) of the $i_\rho$ frequency. This will increase the signal of the coupling between orbital motion and the interfacial mode $i_\rho$ and also its detectability compared to the elastic case. 

\subsection{Distinguishing the twin-star}
The above discussions focus on probing the nuclear phase transition near the surface of the neutron fluid core. Now we discuss the possibility of using the interfacial mode to probe the nuclear-quark phase transition.

As mentioned in Section\,\ref{sec:3}\,B, the $i_{\mu}$-mode frequencies of NS and HS differ up to 7-10\,Hz. In reality, the resonance excitation doesn't happen instantaneously at the resonant orbital frequency. Following \cite{Lai1994}, the total number of orbital cycles that the binary system evolves during the period of the effective resonant excitation of the $i_\mu$ mode is:
\be
\delta N_{{\rm orb,}i}\approx 12\left(\frac{f_i}{100\,{\rm Hz}}\right)^{-5/6}\left(\frac{M}{1.4\,M_{\odot}}\right)^{-5/6}q^{-1/2}\left(\frac{2}{1+q}\right)^{1/6},
\ee
which means that there is a bandwidth of orbital frequency experienced during the resonant excitation. If the internal oscillation frequency of the NS in an NS-NS binary system is differed from that of the HS in an HS-NS binary system with the same mass by $\delta f_i$, and $\delta f_i$ is smaller than the bandwidth of the resonant excitation, this frequency difference is not resolvable and thereby difficult to distinguish the twin-star component.

This can be formulated in another equivalent way. The number of circles for the BNS evolving from $f_1$ to $f_2$ can be obtained by:
\be
\Delta N_{{\rm orb}}=\int_{t_1}^{t_2}\Omega(t)dt/2\pi,
\ee
where the $\Omega(t)$ can be solved from Eq.\,\eqref{eq:orbit_evolution}. The $t_1$ and $t_2$ are the moments when orbital frequencies take the value of $f_1$ and $f_2$, respectively. To distinguish NS/HS from the binary system, $f_{1/2}$ can takes the value of half of the $i_{\mu}$ frequency for NS/HS, that is, the orbital resonant frequency of the NS/HS, respectively. 
The NS/HS is distinguishable as long as the inequality $\Delta N_{{\rm orb}}>\delta N_{{\rm orb}}$ is satisfied.

\begin{figure}[h]
\centering
\includegraphics[width=0.5\textwidth]{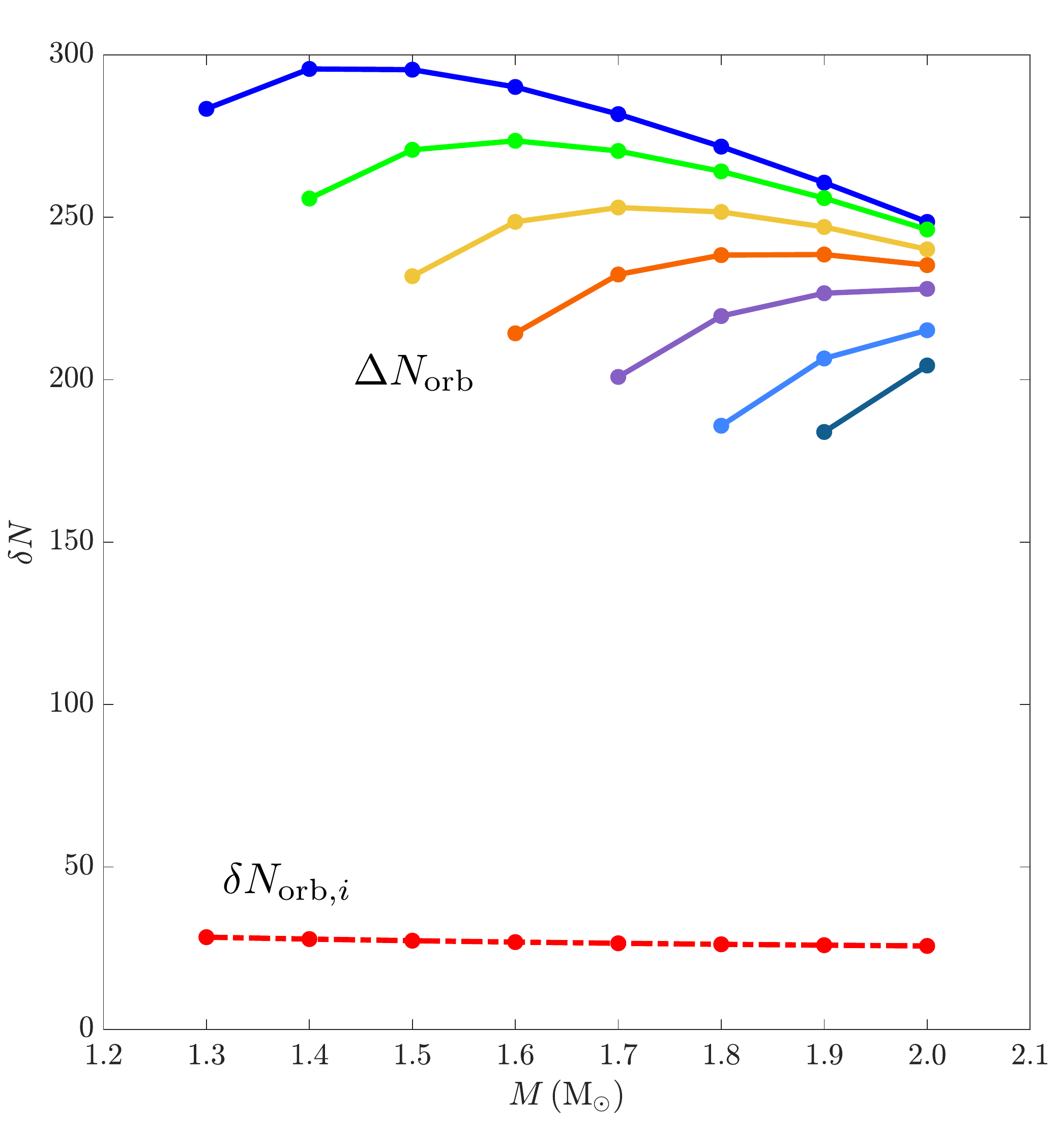}
\caption{Distinguishability of the NS and its HS counterparts at different masses. For illustrative purposes, the mass of the companion compact star is set to be fixed at 1.4\,$M_{\odot}$. The red dashed line is the $\delta N_{{\rm orb}}$ of the $i_{\mu}$-mode resonance of NS. The solid lines are the differences in the orbital cycle between the NS-NS binary system and the NS-HS binary system's resonant orbital frequencies. Detailed analysis is in the main text.}.
\label{fig:twin_star_detectability}
\end{figure}

In Fig.\,\ref{fig:twin_star_detectability}, we demonstrate the distinguishability of the HS solutions compared with the NS solution based on NL3 EoS. Since there is no density discontinuity near the crust-core interface in the NL3 EoS, the corresponding frequency of the crust-core interfacial mode $i_\mu$ is around 30-40\,Hz. Therefore, the orbital excitation of this $i_\mu$-mode happens in a relatively early inspiral stage. Although the frequency differences between NS and HS is only about 7-10\,Hz, the difference of the number of orbital cycles $\Delta N_{{\rm orb}}$ for the HS and NS is large enough compared with the $\delta N_{{\rm orb}}$, which indicates its possible detectability.

\section{Conclusion and Outlook}\label{sec:4}
In this work, we studied the possibility of using the crust-core interfacial mode of the neutron stars to probe the phase transition in dense matter. Some phase transitions, in particular those that can significantly change the mass-radius relation or the tidal deformability of the neutron stars, can be probed using electromagnetic observations or the adiabatic GW waveform. There could also exist phase transitions that have a minor effect on the M-R relation and the tidal deformability. For example, in the fluid core of an NS, first-order phase transitions with density discontinuity can happen near the crust-core interface, due to the possible emergence of the new baryon degrees of freedom. For these phase transitions, we explore the possibility of probing them using the crust-core interfacial mode, which can couple efficiently to the orbital motion of a binary neutron star system and leaves signatures in the gravitational wave radiations. In this work, we carefully analysed the properties of the crust-core interfacial mode and found that they can be significantly modified by the existence of these first-order phase transitions. We also study the observability of this phase transition using Fisher analysis. Moreover, we also explore the effect of the hadron-quark phase transition on the interfacial mode and its possible observability in the gravitational wave signals, as complementary to the measurement of the M-R relation. Our results indicate that a ground-based gravitational wave detector has the potential to detect the signature of these phase transitions via the crust-core interfacial mode.

We want to comment that our results here may have the following inaccuracies. Firstly, our analysis was based on a Newtonian fluid perturbation theory on a general relativistic background stellar structure under the Cowling approximation, where the perturbation of the gravitational field and its coupling to the perturbation of the fluid energy momentum were ignored for simplicity. The most accurate analysis should be based on a fully consistent general relativistic method. For neutron stars, the general relativistic effect is only a small correction that does not change the main results of this work. Secondly, the detailed inelastic mechanical process of the neutron star crust under the excitation of the tidal interactions is not entirely unclear\,\cite{Thompson_2017,Pan2020}, since it depends on the exotic matter phase in the extreme conditions in the neutron star. Therefore the results concerning the inelastic effect of neutron star crust in this work should be valid as semi-quantitative estimations. A more accurate understanding needs a more detailed study of the inelastic properties of the crust.  Thirdly, our analysis is based on the binary system where the compact stars have no spin angular momentum, which in principle can affect the interfacial mode and the orbital motion. The spin affects the internal oscillation mode in a way similar to the Zeeman effect in atomic physics or the Sagnac effect in a laser gyroscope. and the orbital motion of the binary system can also be modulated due to the spin-orbital and spin-spin coupling. Besides, only near-circular binary orbit is considered, while in principle the interfacial mode can also be excited via an eccentric orbit. We leave the improvement of this work by dealing with these inaccuracies in future works.\\

\acknowledgements
The authors devote many thanks to Professor Micaela Oertel, Doctor Helena Pais and the ``CompOSE" team for patient discussions and help in constructing consistent neutron star EoS. Y. M. thanks Professor Huan Yang and Doctor Zhen Pan for the discussions on their work about the interfacial mode, and for sharing their calculations for comparison. He also thanks Professor Shun Wang for his constant support and companionship. J. Z., C. W. and Y. M. thank Professor Yanbei Chen for the discussion and his encouragement in completing this work. We also thank Miss Zeying Xia for her administrative support. J. Z is supported by China National Scholarship (undergraduates). Y. M. is supported by the start-up funding provided by Huazhong University of Science and Technology. E. Z. is supported by National SKA Program of China No. 2020SKA0120300 and NSFC Grant No. 12203017. C. X. is supported by the National SKA Program of China (Grant No. 2020SKA0120300), and the National Natural Science Foundation of China (Grant No. 12275234).

\bibliographystyle{unsrt}
\bibliography{neutron_star}

\begin{thebibliography}{10}

\bibitem{Danielewicz2002_Science298-1592}
Pawe{\l} Danielewicz, Roy Lacey, and William~G. Lynch.
\newblock Determination of the equation of state of dense matter.
\newblock {\em Science}, 298(5598):1592--1596, 2002.

\bibitem{Caplan2017_RMP89-041002}
M.~E. Caplan and C.~J. Horowitz.
\newblock Colloquium: Astromaterial science and nuclear pasta.
\newblock {\em Rev. Mod. Phys.}, 89:041002, Oct 2017.

\bibitem{Baym1971_ApJ170-299}
G.~{Baym}, C.~{Pethick}, and P.~{Sutherland}.
\newblock {The Ground State of Matter at High Densities: Equation of State and
  Stellar Models}.
\newblock {\em Astrophys. J.}, 170:299, December 1971.

\bibitem{Negele1973_NPA207-298}
J.~W. {Negele} and D.~{Vautherin}.
\newblock {Neutron star matter at sub-nuclear densities}.
\newblock {\em Nucl. Phys. A}, 207:298--320, June 1973.

\bibitem{Ravenhall1983_PRL50-2066}
D.~G. Ravenhall, C.~J. Pethick, and J.~R. Wilson.
\newblock Structure of matter below nuclear saturation density.
\newblock {\em Phys. Rev. Lett.}, 50:2066--2069, Jun 1983.

\bibitem{Hashimoto1984_PTP71-320}
Masa-aki Hashimoto, Hironori Seki, and Masami Yamada.
\newblock Shape of nuclei in the crust of neutron star.
\newblock {\em Prog. Theor. Phys.}, 71:320, 1984.

\bibitem{Williams1985_NPA435-844}
R.D. Williams and S.E. Koonin.
\newblock Sub-saturation phases of nuclear matter.
\newblock {\em Nucl. Phys. A}, 435(3):844 -- 858, 1985.

\bibitem{Pethick1998_PLB427-7}
C.J. Pethick and A.Y. Potekhin.
\newblock Liquid crystals in the mantles of neutron stars.
\newblock {\em Phys. Lett. B}, 427(1):7 -- 12, 1998.

\bibitem{Oyamatsu1993_NPA561-431}
K.~Oyamatsu.
\newblock Nuclear shapes in the inner crust of a neutron star.
\newblock {\em Nucl. Phys. A}, 561(3):431 -- 452, 1993.

\bibitem{Maruyama2005_PRC72-015802}
Toshiki Maruyama, Toshitaka Tatsumi, Dmitri~N. Voskresensky, Tomonori Tanigawa,
  and Satoshi Chiba.
\newblock Nuclear ``pasta'' structures and the charge screening effect.
\newblock {\em Phys. Rev. C}, 72:015802, Jul 2005.

\bibitem{Togashi2017_NPA961-78}
H.~Togashi, K.~Nakazato, Y.~Takehara, S.~Yamamuro, H.~Suzuki, and M.~Takano.
\newblock Nuclear equation of state for core-collapse supernova simulations
  with realistic nuclear forces.
\newblock {\em Nucl. Phys. A}, 961:78 -- 105, 2017.

\bibitem{Shen2011_ApJ197-20}
H.~Shen, H.~Toki, K.~Oyamatsu, and K.~Sumiyoshi.
\newblock Relativistic equation of state for core-collapse supernova
  simulations.
\newblock {\em Astrophys. J.}, 197(2):20, 2011.

\bibitem{Demorest2010_Nature467-1081}
P.~B. Demorest, T.~Pennucci, S.~M. Ransom, M.~S.~E. Roberts, and J.~W.~T.
  Hessels.
\newblock A two-solar-mass neutron star measured using shapiro delay.
\newblock {\em Nature}, 467:1081--1083, 2010.

\bibitem{Antoniadis2013_Science340-1233232}
John Antoniadis, Paulo C.~C. Freire, Norbert Wex, Thomas~M. Tauris, Ryan~S.
  Lynch, Marten~H. van Kerkwijk, Michael Kramer, Cees Bassa, Vik~S. Dhillon,
  Thomas Driebe, Jason W.~T. Hessels, Victoria~M. Kaspi, Vladislav~I.
  Kondratiev, Norbert Langer, Thomas~R. Marsh, Maura~A. McLaughlin, Timothy~T.
  Pennucci, Scott~M. Ransom, Ingrid~H. Stairs, Joeri van Leeuwen, Joris P.~W.
  Verbiest, and David~G. Whelan.
\newblock A massive pulsar in a compact relativistic binary.
\newblock {\em Science}, 340:1233232, 2013.

\bibitem{Fonseca2016_ApJ832-167}
Emmanuel Fonseca, Timothy~T. Pennucci, Justin~A. Ellis, Ingrid~H. Stairs,
  David~J. Nice, Scott~M. Ransom, Paul~B. Demorest, Zaven Arzoumanian, Kathryn
  Crowter, Timothy Dolch, Robert~D. Ferdman, Marjorie~E. Gonzalez, Glenn Jones,
  Megan~L. Jones, Michael~T. Lam, Lina Levin, Maura~A. McLaughlin, Kevin
  Stovall, Joseph~K. Swiggum, and Weiwei Zhu.
\newblock The nanograv nine-year data set: Mass and geometric measurements of
  binary millisecond pulsars.
\newblock {\em Astrophys. J.}, 832(2):167, 2016.

\bibitem{Cromartie2020_NA4-72}
H.~Thankful Cromartie, Emmanuel Fonseca, Scott~M. Ransom, Paul~B. Demorest,
  Zaven Arzoumanian, Harsha Blumer, Paul~R. Brook, Megan~E. DeCesar, Timothy
  Dolch, Justin~A. Ellis, Robert~D. Ferdman, Elizabeth~C. Ferrara, Nathaniel
  Garver-Daniels, Pete~A. Gentile, Megan~L. Jones, Michael~T. Lam, Duncan~R.
  Lorimer, Ryan~S. Lynch, Maura~A. McLaughlin, Cherry Ng, David~J. Nice,
  Timothy~T. Pennucci, Renee Spiewak, Ingrid~H. Stairs, Kevin Stovall,
  Joseph~K. Swiggum, and Weiwei Zhu.
\newblock Relativistic shapiro delay measurements of an extremely massive
  millisecond pulsar.
\newblock {\em Nat. Astron.}, 4:72--76, 2020.

\bibitem{Fonseca2021_ApJ915-L12}
E.~Fonseca, H.~T. Cromartie, T.~T. Pennucci, P.~S. Ray, A.~Yu. Kirichenko,
  S.~M. Ransom, P.~B. Demorest, I.~H. Stairs, Z.~Arzoumanian, L.~Guillemot,
  A.~Parthasarathy, M.~Kerr, I.~Cognard, P.~T. Baker, H.~Blumer, P.~R. Brook,
  M.~DeCesar, T.~Dolch, F.~A. Dong, E.~C. Ferrara, W.~Fiore, N.~Garver-Daniels,
  D.~C. Good, R.~Jennings, M.~L. Jones, V.~M. Kaspi, M.~T. Lam, D.~R. Lorimer,
  J.~Luo, A.~McEwen, J.~W. McKee, M.~A. McLaughlin, N.~McMann, B.~W. Meyers,
  A.~Naidu, C.~Ng, D.~J. Nice, N.~Pol, H.~A. Radovan, B.~Shapiro-Albert, C.~M.
  Tan, S.~P. Tendulkar, J.~K. Swiggum, H.~M. Wahl, and W.~W. Zhu.
\newblock Refined mass and geometric measurements of the high-mass {PSR}
  j0740+6620.
\newblock {\em Astrophys. J.}, 915(1):L12, jul 2021.

\bibitem{LVC2018_PRL121-161101}
{LIGO Scientific and Virgo Collaborations}.
\newblock Gw170817: Measurements of neutron star radii and equation of state.
\newblock {\em Phys. Rev. Lett.}, 121(16):161101, 2018.

\bibitem{Riley2019_ApJ887-L21}
T.~E. Riley, A.~L. Watts, S.~Bogdanov, P.~S. Ray, R.~M. Ludlam, S.~Guillot,
  Z.~Arzoumanian, C.~L. Baker, A.~V. Bilous, D.~Chakrabarty, K.~C. Gendreau,
  A.~K. Harding, W.~C.~G. Ho, J.~M. Lattimer, S.~M. Morsink, and T.~E.
  Strohmayer.
\newblock A nicer view of psr j0030+0451: Millisecond pulsar parameter
  estimation.
\newblock {\em Astrophys. J.}, 887(1):L21, dec 2019.

\bibitem{Riley2021_ApJ918-L27}
Thomas~E. Riley, Anna~L. Watts, Paul~S. Ray, Slavko Bogdanov, Sebastien
  Guillot, Sharon~M. Morsink, Anna~V. Bilous, Zaven Arzoumanian, Devarshi
  Choudhury, Julia~S. Deneva, Keith~C. Gendreau, Alice~K. Harding, Wynn C.~G.
  Ho, James~M. Lattimer, Michael Loewenstein, Renee~M. Ludlam, Craig~B.
  Markwardt, Takashi Okajima, Chanda Prescod-Weinstein, Ronald~A. Remillard,
  Michael~T. Wolff, Emmanuel Fonseca, H.~Thankful Cromartie, Matthew Kerr,
  Timothy~T. Pennucci, Aditya Parthasarathy, Scott Ransom, Ingrid Stairs, Lucas
  Guillemot, and Ismael Cognard.
\newblock A {NICER} view of the massive pulsar {PSR} j0740+6620 informed by
  radio timing and {XMM}-newton spectroscopy.
\newblock {\em Astrophys. J.}, 918(2):L27, sep 2021.

\bibitem{Miller2019_ApJ887-L24}
M.~C. Miller, F.~K. Lamb, A.~J. Dittmann, S.~Bogdanov, Z.~Arzoumanian, K.~C.
  Gendreau, S.~Guillot, A.~K. Harding, W.~C.~G. Ho, J.~M. Lattimer, R.~M.
  Ludlam, S.~Mahmoodifar, S.~M. Morsink, P.~S. Ray, T.~E. Strohmayer, K.~S.
  Wood, T.~Enoto, R.~Foster, T.~Okajima, G.~Prigozhin, and Y.~Soong.
\newblock Psr j0030+0451 mass and radius from nicer data and implications for
  the properties of neutron star matter.
\newblock {\em Astrophys. J.}, 887(1):L24, dec 2019.

\bibitem{Miller2021_ApJ918-L28}
M.~C. Miller, F.~K. Lamb, A.~J. Dittmann, S.~Bogdanov, Z.~Arzoumanian, K.~C.
  Gendreau, S.~Guillot, W.~C.~G. Ho, J.~M. Lattimer, M.~Loewenstein, S.~M.
  Morsink, P.~S. Ray, M.~T. Wolff, C.~L. Baker, T.~Cazeau, S.~Manthripragada,
  C.~B. Markwardt, T.~Okajima, S.~Pollard, I.~Cognard, H.~T. Cromartie,
  E.~Fonseca, L.~Guillemot, M.~Kerr, A.~Parthasarathy, T.~T. Pennucci,
  S.~Ransom, and I.~Stairs.
\newblock The radius of {PSR} j0740+6620 from {NICER} and {XMM}-newton data.
\newblock {\em Astrophys. J.}, 918(2):L28, sep 2021.

\bibitem{Abbott2018tidal}
B.~P. {Abbott}, R.~{Abbott}, T.~D. {Abbott}, F.~{Acernese}, K.~{Ackley},
  C.~{Adams}, T.~{Adams}, P.~{Addesso}, R.~X. {Adhikari}, V.~B. {Adya}, and
  et~al.
\newblock {GW170817: Measurements of Neutron Star Radii and Equation of State}.
\newblock {\em Physical Review Letters}, 121(16):161101, October 2018.

\bibitem{Abbottprx2019}
B.~P. {Abbott}, R.~{Abbott}, T.~D. {Abbott}, F.~{Acernese}, K.~{Ackley},
  C.~{Adams}, T.~{Adams}, P.~{Addesso}, R.~X. {Adhikari}, V.~B. {Adya}, and
  et~al.
\newblock Properties of the binary neutron star merger gw170817.
\newblock {\em Phys. Rev. X}, 9:011001, Jan 2019.

\bibitem{Annala2017}
E.~{Annala}, T.~{Gorda}, A.~{Kurkela}, and A.~{Vuorinen}.
\newblock {Gravitational-Wave Constraints on the Neutron-Star-Matter Equation
  of State}.
\newblock {\em Physical Review Letters}, 120(17):172703, April 2018.

\bibitem{De2018}
Soumi {De}, Daniel {Finstad}, James~M. {Lattimer}, Duncan~A. {Brown}, Edo
  {Berger}, and Christopher~M. {Biwer}.
\newblock {Tidal Deformabilities and Radii of Neutron Stars from the
  Observation of GW170817}.
\newblock {\em \prl}, 121(9):091102, Aug 2018.

\bibitem{Ruiz2017}
M.~{Ruiz}, S.~L. {Shapiro}, and A.~{Tsokaros}.
\newblock {GW170817, general relativistic magnetohydrodynamic simulations, and
  the neutron star maximum mass}.
\newblock {\em Phys. Rev. D}, 97(2):021501, January 2018.

\bibitem{Rezzolla2017}
L.~{Rezzolla}, E.~R. {Most}, and L.~R. {Weih}.
\newblock {Using Gravitational-wave Observations and Quasi-universal Relations
  to Constrain the Maximum Mass of Neutron Stars}.
\newblock {\em Astrophys. J. Lett.}, 852:L25, January 2018.

\bibitem{Shibata2019}
Masaru {Shibata}, Enping {Zhou}, Kenta {Kiuchi}, and Sho {Fujibayashi}.
\newblock {Constraint on the maximum mass of neutron stars using GW170817
  event}.
\newblock {\em \prd}, 100(2):023015, Jul 2019.

\bibitem{Margalit2017}
B.~{Margalit} and B.~D. {Metzger}.
\newblock {Constraining the Maximum Mass of Neutron Stars from Multi-messenger
  Observations of GW170817}.
\newblock {\em Astrophys. J. Lett.}, 850:L19, December 2017.

\bibitem{Bauswein2017b}
A.~{Bauswein}, O.~{Just}, H.-T. {Janka}, and N.~{Stergioulas}.
\newblock {Neutron-star Radius Constraints from GW170817 and Future
  Detections}.
\newblock {\em Astrophys. J. Lett.}, 850:L34, December 2017.

\bibitem{Lau2021}
Shu~Yan Lau and Kent Yagi.
\newblock Probing hybrid stars with gravitational waves via interfacial modes.
\newblock {\em Phys. Rev. D}, 103:063015, Mar 2021.

\bibitem{McDermott1985}
P.~N. {McDermott}, H.~M. {van Horn}, and C.~J. {Hansen}.
\newblock {Nonradial Oscillations of Neutron Stars}.
\newblock {\em \apj}, 325:725, February 1988.

\bibitem{Tsang2012}
David Tsang, Jocelyn~S. Read, Tanja Hinderer, Anthony~L. Piro, and Ruxandra
  Bondarescu.
\newblock Resonant shattering of neutron star crusts.
\newblock {\em Phys. Rev. Lett.}, 108:011102, Jan 2012.

\bibitem{Pan2020}
Zhen Pan, Zhenwei Lyu, B\'eatrice Bonga, N\'estor Ortiz, and Huan Yang.
\newblock Probing crust meltdown in inspiraling binary neutron stars.
\newblock {\em Phys. Rev. Lett.}, 125:201102, Nov 2020.

\bibitem{Alford2013}
Mark~G. Alford, Sophia Han, and Madappa Prakash.
\newblock Generic conditions for stable hybrid stars.
\newblock {\em Phys. Rev. D}, 88:083013, Oct 2013.

\bibitem{Kruger2015}
C.~J. Kr\"uger, W.~C.~G. Ho, and N.~Andersson.
\newblock Seismology of adolescent neutron stars: Accounting for thermal
  effects and crust elasticity.
\newblock {\em Phys. Rev. D}, 92:063009, Sep 2015.

\bibitem{Miao2020_ApJ904-103}
Zhiqiang Miao, Ang Li, Zhenyu Zhu, and Sophia Han.
\newblock Constraining hadron-quark phase transition parameters within the
  quark-mean-field model using multimessenger observations of neutron stars.
\newblock {\em Astrophys. J.}, 904(2):103, nov 2020.

\bibitem{Kurkela_2010}
Aleksi Kurkela, Paul Romatschke, Aleksi Vuorinen, and Bin Wu.
\newblock Looking inside neutron stars: Microscopic calculations confront
  observations, 2010.

\bibitem{Alford_2015}
Mark~G. Alford, G.~F. Burgio, S.~Han, G.~Taranto, and D.~Zappal\`a.
\newblock Constraining and applying a generic high-density equation of state.
\newblock {\em Phys. Rev. D}, 92:083002, Oct 2015.

\bibitem{Baym_2018}
Gordon Baym, Tetsuo Hatsuda, Toru Kojo, Philip~D Powell, Yifan Song, and
  Tatsuyuki Takatsuka.
\newblock From hadrons to quarks in neutron stars: a review.
\newblock {\em Reports on Progress in Physics}, 81(5):056902, mar 2018.

\bibitem{BAYM1976241}
G.~Baym and S.A. Chin.
\newblock Can a neutron star be a giant mit bag?
\newblock {\em Physics Letters B}, 62(2):241--244, 1976.

\bibitem{Fortin2016}
M.~Fortin, C.~Provid\^encia, Ad.~R. Raduta, F.~Gulminelli, J.~L. Zdunik,
  P.~Haensel, and M.~Bejger.
\newblock Neutron star radii and crusts: Uncertainties and unified equations of
  state.
\newblock {\em Phys. Rev. C}, 94:035804, Sep 2016.

\bibitem{Lalazissis1997_PRC55-540}
G.~A. Lalazissis, J.~K\"onig, and P.~Ring.
\newblock New parametrization for the lagrangian density of relativistic mean
  field theory.
\newblock {\em Phys. Rev. C}, 55:540--543, Jan 1997.

\bibitem{Brush1966_JCP45-2102}
S.~G. Brush, H.~L. Sahlin, and E.~Teller.
\newblock Monte carlo study of a one-component plasma. i.
\newblock {\em J. Chem. Phys.}, 45(6):2102--2118, 1966.

\bibitem{Ogata1993_ApJ417-265}
Shuji {Ogata}, Setsuo {Ichimaru}, and Hugh~M. {van Horn}.
\newblock {Thermonuclear Reaction Rates for Dense Binary-Ionic Mixtures}.
\newblock {\em Astrophys. J.}, 417:265, November 1993.

\bibitem{Jones1996_PRL76-4572}
M.~D. Jones and D.~M. Ceperley.
\newblock Crystallization of the one-component plasma at finite temperature.
\newblock {\em Phys. Rev. Lett.}, 76:4572--4575, Jun 1996.

\bibitem{Potekhin2000_PRE62-8554}
Alexander~Y. Potekhin and Gilles Chabrier.
\newblock Equation of state of fully ionized electron-ion plasmas. ii.
  extension to relativistic densities and to the solid phase.
\newblock {\em Phys. Rev. E}, 62:8554--8563, Dec 2000.

\bibitem{Medin2010_PRE81-036107}
Zach Medin and Andrew Cumming.
\newblock Crystallization of classical multicomponent plasmas.
\newblock {\em Phys. Rev. E}, 81:036107, Mar 2010.

\bibitem{Caplan2018_ApJ860-148}
M.~E. Caplan, Andrew Cumming, D.~K. Berry, C.~J. Horowitz, and R.~Mckinven.
\newblock Polycrystalline crusts in accreting neutron stars.
\newblock {\em Astrophys. J.}, 860(2):148, jun 2018.

\bibitem{Baym2018_RPP81-056902}
Gordon Baym, Tetsuo Hatsuda, Toru Kojo, Philip~D Powell, Yifan Song, and
  Tatsuyuki Takatsuka.
\newblock From hadrons to quarks in neutron stars: a review.
\newblock {\em Rep. Prog. Phys.}, 81(5):056902, 2018.

\bibitem{Sun2019_PRD99-023004}
Ting-Ting Sun, Shi-Sheng Zhang, Qiu-Lan Zhang, and Cheng-Jun Xia.
\newblock Strangeness and $\mathrm{\ensuremath{\Delta}}$ resonance in compact
  stars with relativistic-mean-field models.
\newblock {\em Phys. Rev. D}, 99:023004, Jan 2019.

\bibitem{Annala2020_NP}
Eemeli Annala, Tyler Gorda, Aleksi Kurkela, Joonas N\"attil\"a, and Aleksi
  Vuorinen.
\newblock Evidence for quark-matter cores in massive neutron stars.
\newblock {\em Nat. Phys.}, 16:907, 2020.

\bibitem{Dexheimer2021_PRC103-025808}
V.~Dexheimer, R.~O. Gomes, T.~Kl\"ahn, S.~Han, and M.~Salinas.
\newblock Gw190814 as a massive rapidly rotating neutron star with exotic
  degrees of freedom.
\newblock {\em Phys. Rev. C}, 103:025808, Feb 2021.

\bibitem{Tan2022_PRD105-023018}
Hung Tan, Travis Dore, Veronica Dexheimer, Jacquelyn Noronha-Hostler, and
  Nicol\'as Yunes.
\newblock Extreme matter meets extreme gravity: Ultraheavy neutron stars with
  phase transitions.
\newblock {\em Phys. Rev. D}, 105:023018, Jan 2022.

\bibitem{Akmal1998_PRC58-1804}
A.~Akmal, V.~R. Pandharipande, and D.~G. Ravenhall.
\newblock Equation of state of nucleon matter and neutron star structure.
\newblock {\em Phys. Rev. C}, 58:1804--1828, Sep 1998.

\bibitem{Xie2021_PRC103-035802}
Wen-Jie Xie and Bao-An Li.
\newblock Bayesian inference of the dense-matter equation of state
  encapsulating a first-order hadron-quark phase transition from observables of
  canonical neutron stars.
\newblock {\em Phys. Rev. C}, 103:035802, Mar 2021.

\bibitem{Jin2022_PLB829-137121}
Hao-Miao Jin, Cheng-Jun Xia, Ting-Ting Sun, and Guang-Xiong Peng.
\newblock Quark condensate and chiral symmetry restoration in neutron stars.
\newblock {\em Phys. Lett. B}, 829:137121, 2022.

\bibitem{Pfaff2022_PRC105-035802}
Antoine Pfaff, Hubert Hansen, and Francesca Gulminelli.
\newblock Bayesian analysis of the properties of hybrid stars with the
  nambu--jona-lasinio model.
\newblock {\em Phys. Rev. C}, 105:035802, Mar 2022.

\bibitem{McDermott1988_ApJ325-725}
P.~N. {McDermott}, H.~M. {van Horn}, and C.~J. {Hansen}.
\newblock {Nonradial Oscillations of Neutron Stars}.
\newblock {\em Astrophys. J.}, 325:725, February 1988.

\bibitem{Schumaker1983_MNRAS203-457}
Bonny~L. Schumaker and Kip~S. Thorne.
\newblock Torsional oscillations of neutron stars.
\newblock {\em Mon. Not. R. Astron. Soc.}, 203(2):457--489, 06 1983.

\bibitem{Hansen1980_ApJ238-740}
C.~J. {Hansen} and D.~F. {Cioffi}.
\newblock {Torsional oscillations in neutron star crusts}.
\newblock {\em Astrophys. J.}, 238:740--742, June 1980.

\bibitem{Passamonti2012_MNRAS419-638}
A.~Passamonti and N.~Andersson.
\newblock Towards real neutron star seismology: accounting for elasticity and
  superfluidity.
\newblock {\em Mon. Not. R. Astron. Soc.}, 419(1):638--655, 2012.

\bibitem{Passamonti2021}
A.~{Passamonti}, N.~{Andersson}, and P.~{Pnigouras}.
\newblock {Dynamical tides in neutron stars: the impact of the crust}.
\newblock {\em Mon. Not. R. Astron. Soc.}, 504(1):1273--1293, June 2021.

\bibitem{2015PPN....46..633T}
S.~{Typel}, M.~{Oertel}, and T.~{Kl{\"a}hn}.
\newblock {CompOSE CompStar online supernova equations of state harmonising the
  concert of nuclear physics and astrophysics compose.obspm.fr}.
\newblock {\em Physics of Particles and Nuclei}, 46(4):633--664, July 2015.

\bibitem{2017RvMP...89a5007O}
M.~{Oertel}, M.~{Hempel}, T.~{Kl{\"a}hn}, and S.~{Typel}.
\newblock {Equations of state for supernovae and compact stars}.
\newblock {\em Reviews of Modern Physics}, 89(1):015007, January 2017.

\bibitem{https://doi.org/10.48550/arxiv.2203.03209}
S.~Typel, M.~Oertel, T.~Klähn, D.~Chatterjee, V.~Dexheimer, C.~Ishizuka,
  M.~Mancini, J.~Novak, H.~Pais, C.~Providencia, A.~Raduta, M.~Servillat, and
  L.~Tolos.
\newblock Compose reference manual, 2022.

\bibitem{Malik_2022}
Tuhin Malik and Helena Pais.
\newblock Inner crust equations of state for {CompOSE}.
\newblock {\em The European Physical Journal A}, 58(8), aug 2022.

\bibitem{REINHARD1995467}
P.-G. Reinhard and H.~Flocard.
\newblock Nuclear effective forces and isotope shifts.
\newblock {\em Nuclear Physics A}, 584(3):467--488, 1995.

\bibitem{PhysRevC.33.335}
J.~Friedrich and P.-G. Reinhard.
\newblock Skyrme-force parametrization: Least-squares fit to nuclear
  ground-state properties.
\newblock {\em Phys. Rev. C}, 33:335--351, Jan 1986.

\bibitem{CHABANAT1998231}
E.~Chabanat, P.~Bonche, P.~Haensel, J.~Meyer, and R.~Schaeffer.
\newblock A skyrme parametrization from subnuclear to neutron star densities
  part ii. nuclei far from stabilities.
\newblock {\em Nuclear Physics A}, 635(1):231--256, 1998.

\bibitem{PhysRevC.58.1804}
A.~Akmal, V.~R. Pandharipande, and D.~G. Ravenhall.
\newblock Equation of state of nucleon matter and neutron star structure.
\newblock {\em Phys. Rev. C}, 58:1804--1828, Sep 1998.

\bibitem{PhysRevC.81.015803}
S.~Typel, G.~R\"opke, T.~Kl\"ahn, D.~Blaschke, and H.~H. Wolter.
\newblock Composition and thermodynamics of nuclear matter with light clusters.
\newblock {\em Phys. Rev. C}, 81:015803, Jan 2010.

\bibitem{PhysRevC.90.045803}
Fabrizio Grill, Helena Pais, Constan\ifmmode \mbox{\c{c}}\else~\c{c}\fi{}a
  Provid\^encia, Isaac Vida\~na, and Sidney~S. Avancini.
\newblock Equation of state and thickness of the inner crust of neutron stars.
\newblock {\em Phys. Rev. C}, 90:045803, Oct 2014.

\bibitem{PhysRevC.71.024312}
G.~A. Lalazissis, T.~Nik\ifmmode \check{s}\else
  \v{s}\fi{}i\ifmmode~\acute{c}\else \'{c}\fi{}, D.~Vretenar, and P.~Ring.
\newblock New relativistic mean-field interaction with density-dependent
  meson-nucleon couplings.
\newblock {\em Phys. Rev. C}, 71:024312, Feb 2005.

\bibitem{Xia_2022}
Cheng-Jun Xia, Toshiki Maruyama, Ang Li, Bao~Yuan Sun, Wen-Hui Long, and
  Ying-Xun Zhang.
\newblock Unified neutron star eoss and neutron star structures in rmf models.
\newblock {\em Communications in Theoretical Physics}, 74(9):095303, aug 2022.

\bibitem{PhysRevC.105.045803}
Cheng-Jun Xia, Bao~Yuan Sun, Toshiki Maruyama, Wen-Hui Long, and Ang Li.
\newblock Unified nuclear matter equations of state constrained by the
  in-medium balance in density-dependent covariant density functionals.
\newblock {\em Phys. Rev. C}, 105:045803, Apr 2022.

\bibitem{TYPEL1999331}
S.~Typel and H.H. Wolter.
\newblock Relativistic mean field calculations with density-dependent
  meson-nucleon coupling.
\newblock {\em Nuclear Physics A}, 656(3):331--364, 1999.

\bibitem{Wei_2020}
Bin Wei, Qiang Zhao, Zhi-Heng Wang, Jing Geng, Bao-Yuan Sun, Yi-Fei Niu, and
  Wen-Hui Long.
\newblock Novel relativistic mean field lagrangian guided by pseudo-spin
  symmetry restoration *.
\newblock {\em Chinese Physics C}, 44(7):074107, jul 2020.

\bibitem{Flanagan2007}
\'Eanna~\'E. Flanagan and \'Etienne Racine.
\newblock Gravitomagnetic resonant excitation of rossby modes in coalescing
  neutron star binaries.
\newblock {\em Phys. Rev. D}, 75:044001, Feb 2007.

\bibitem{Sizheng2021}
Sizheng Ma, Hang Yu, and Yanbei Chen.
\newblock Detecting resonant tidal excitations of rossby modes in coalescing
  neutron-star binaries with third-generation gravitational-wave detectors.
\newblock {\em Phys. Rev. D}, 103:063020, Mar 2021.

\bibitem{Lai1994}
Dong Lai.
\newblock {Resonant oscillations and tidal heating in coalescing binary neutron
  stars}.
\newblock {\em Monthly Notices of the Royal Astronomical Society},
  270(3):611--629, 10 1994.

\bibitem{Cutler1994}
Curt {Cutler} and {\'E}anna~E. {Flanagan}.
\newblock {Gravitational waves from merging compact binaries: How accurately
  can one extract the binary's parameters from the inspiral
  waveform\textbackslash?}
\newblock {\em \prd}, 49(6):2658--2697, March 1994.

\bibitem{Yu2017}
Hang {Yu} and Nevin~N. {Weinberg}.
\newblock {Resonant tidal excitation of superfluid neutron stars in coalescing
  binaries}.
\newblock {\em Mon. Not. R. Astron. Soc.}, 464(3):2622--2637, January 2017.

\bibitem{Aasi_2015}
The LIGO~Scientific Collaboration.
\newblock Advanced ligo.
\newblock {\em Classical and Quantum Gravity}, 32(7):074001, mar 2015.

\bibitem{Dwyer2015}
Sheila Dwyer, Daniel Sigg, Stefan~W. Ballmer, Lisa Barsotti, Nergis Mavalvala,
  and Matthew Evans.
\newblock Gravitational wave detector with cosmological reach.
\newblock {\em Phys. Rev. D}, 91:082001, Apr 2015.

\bibitem{Thompson_2017}
Christopher Thompson, Huan Yang, and Néstor Ortiz.
\newblock Global crustal dynamics of magnetars in relation to their bright
  x-ray outbursts.
\newblock {\em The Astrophysical Journal}, 841(1):54, may 2017.

\bibitem{1993ApJ...414..695C}
Gilles {Chabrier}.
\newblock {Quantum Effects in Dense Coulombic Matter: Application to the
  Cooling of White Dwarfs}.
\newblock {\em \apj}, 414:695, September 1993.

\end{thebibliography}

\appendix
\section{Some details on solving the interfacial modes}
We follow the standard approach\,\cite{McDermott1988_ApJ325-725} to solve the eigenfrequencies of asteroseismology under the Cowling approximation, briefly summarised in this section. We also discuss some important numerical details.

In the solid crust, the movement of a mass element is governed by the Newton's second law, mass continuity equation and the Poisson equation for gravitational potential: 
\be
\begin{split}
&\frac{\partial \vec{v}}{\partial t}+(\vec{v}\cdot \nabla)\vec{v}=\frac{1}{\rho}\nabla \cdot \mathbb{S}-\nabla \Phi,\\
&\frac{\partial \rho}{\partial t}+\nabla \cdot (\rho \vec{v})=0,\quad \nabla^2 \Phi = 4 \pi G \rho.
\end{split}
\ee
where $\mathbb{S}$ is the elastic stress tensor with component as:
\be
S_{ij}=\Gamma_1 P {\rm Tr}(\epsilon) \delta_{ij}+2\mu [\epsilon_{ij}-\frac{1}{3}{\rm Tr}(\epsilon)\delta_{ij}]
\ee
where $\epsilon_{ij}=(\xi_{i,j}+\xi_{j,i})/2$ is the strain tensor, $\mu$ is the shear modulus and $\Gamma_1=dlnP/dln\rho$ is the adiabatic index, which is a characterisation of the matter hardness. In particular, we will see later that the eigen-mode results are very sensitive to $\Gamma$ and we need to be careful in the numerical treatment of $\Gamma$. For the fluid core, the equation is the same except the shear modulus $\mu$ is set to be zero. 

Perturbations are applied to the above equations, with static zeroth order background quantities obtained using Tolman-Oppenheimer-Volkoff  equation. We do not list the cumbersome detailed form of the perturbation equations, which can be easily found in\,\cite{McDermott1985,McDermott1988_ApJ325-725,Pan2020,Lau2021}, we only list the important junction conditions at different interfaces, and other important details that needs to be noticed.

$\bullet$\,\textbf{Junction conditions}---
As mentioned in Sec\,\ref{sec:3}, there are two different interfaces in the compact star: the shear-modulus discontinuity interface and the density discontinuity interface. When integrating across these interfaces, proper junction conditions must be applied to connect the perturbation variables at both sides of the interface.

The physical conditions which must be satisfied at both interfaces are the continuity of the radial displacement of the fluid elements and the radial stress. These two requirements is sufficient to construct the connection at the density discontinuity interface. At the shear-modulus interface, an additional condition acquiring zero shear stress means that an ideal fluid with $\mu=0$ does not support shear stress. 


The dimensionless perturbation variables of the solid crust that appear in the oscillation equations are:  
\be
\begin{split}
  z_1&=\frac{U}{r},\,\,\,\,\,\,z_2=\frac{\lambda}{p} [\frac{1}{r^2}\frac{d}{dr}(r^2U)-\frac{l(l+1)}{r}V)]+2\frac{\mu}{p}\frac{dU}{dr},\\
  z_3&=\frac{V}{r},\,\,\,\,\,\,z_4=\frac{\mu}{p}(\frac{dV}{dr}-\frac{V}{r}+\frac{U}{r}),
\end{split}
\ee
Where $\lambda=\Gamma_1p-2/3\mu$ is the Lam\'e coefficient.

The dimensionless perturbation variables of the liquid core are:
\be
y_1=\frac{U}{r},\,\,\,\,\,\,y_2=\frac{\delta p}{\rho gr}.
\ee
Correspondingly, the junction conditions at the two interfaces are:
\be
\begin{split}
\label{eq: crust-liquid junction}
     [z_1]_{{\rm solid}}&=[y_1]_{{\rm liquid}}\, \\
    [z_2]_{{\rm solid}}&=[\tilde{V}(y_1-y_2)]_{{\rm liquid}}\, \\
    [z_4]_{{\rm solid}}&=0\,.
\end{split}
\ee
and
\be
\begin{split}
\label{eq: rho interface junction}
    y_{1l}&=y_{1h}\,,\\
    \tilde{V}_l(y_{1l}-y_{2l})&=\tilde{V}_h(y_{1h}-y_{2h})\,.
\end{split}
\ee
In the junction condition, there appears a quantity $\tilde{V}=-d\,{\rm ln} p/d\,{\rm ln} r=\rho g r/p$. Therefore the density discontinuity will results in a discontinuity of $\tilde{V}$. 

The brackets "[\,]" in the above formula with the subscripts "solid" and "liquid" indicates that the quantities should be evaluated at both solid and liquid sides respectively.  In\,\eqref{eq: rho interface junction}, the subscripts "$l$" and "$h$" indicates that the quantities should be evaluated at the lower mass density side and the higher mass density side respectively at the interface of the first order phase transition in the liquid core, which is essential to the calculation of "free-slipping" effects. 

\begin{figure}[h]
\centering
\includegraphics[width=0.45\textwidth]{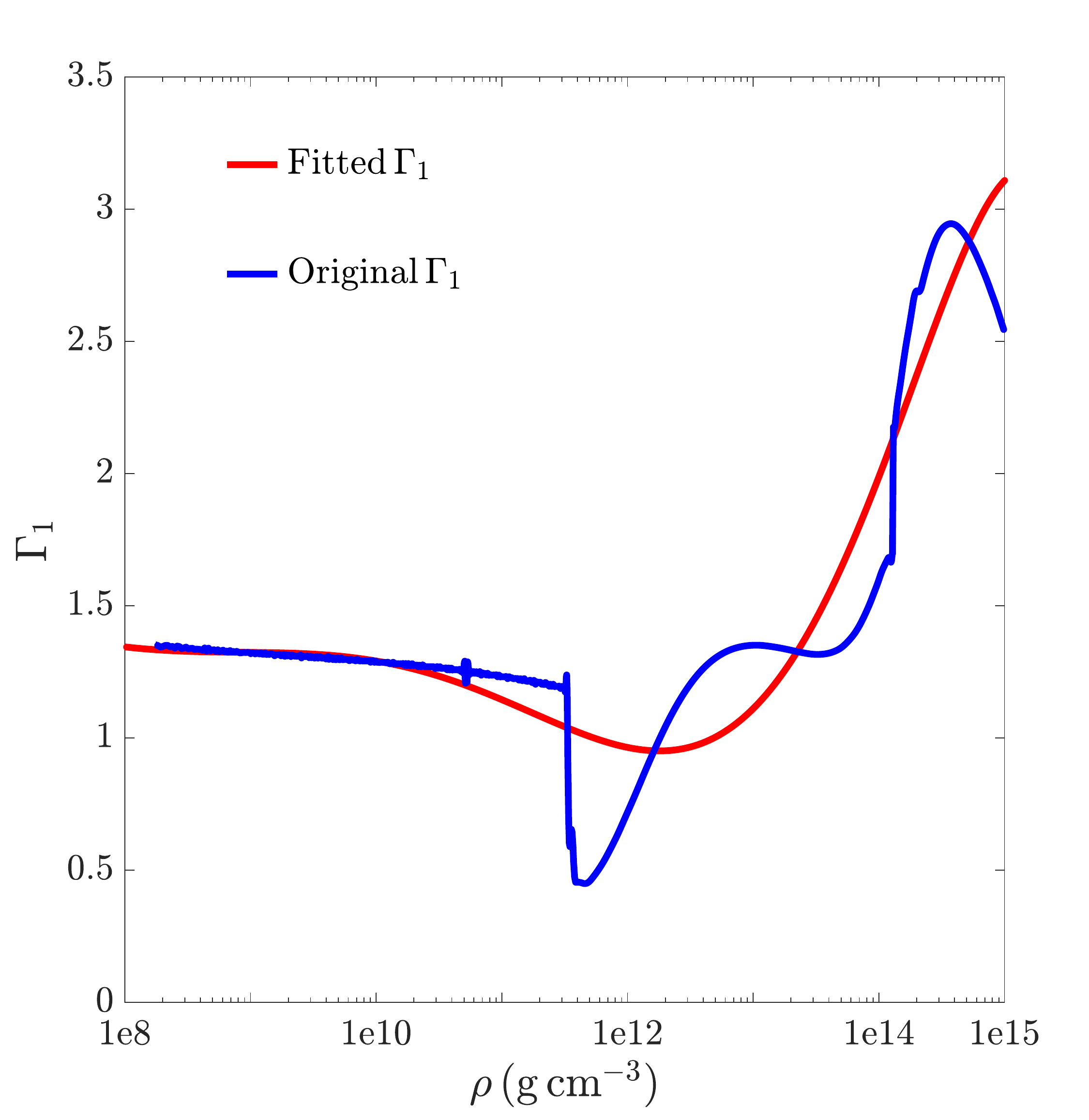}
\includegraphics[width=0.45\textwidth]{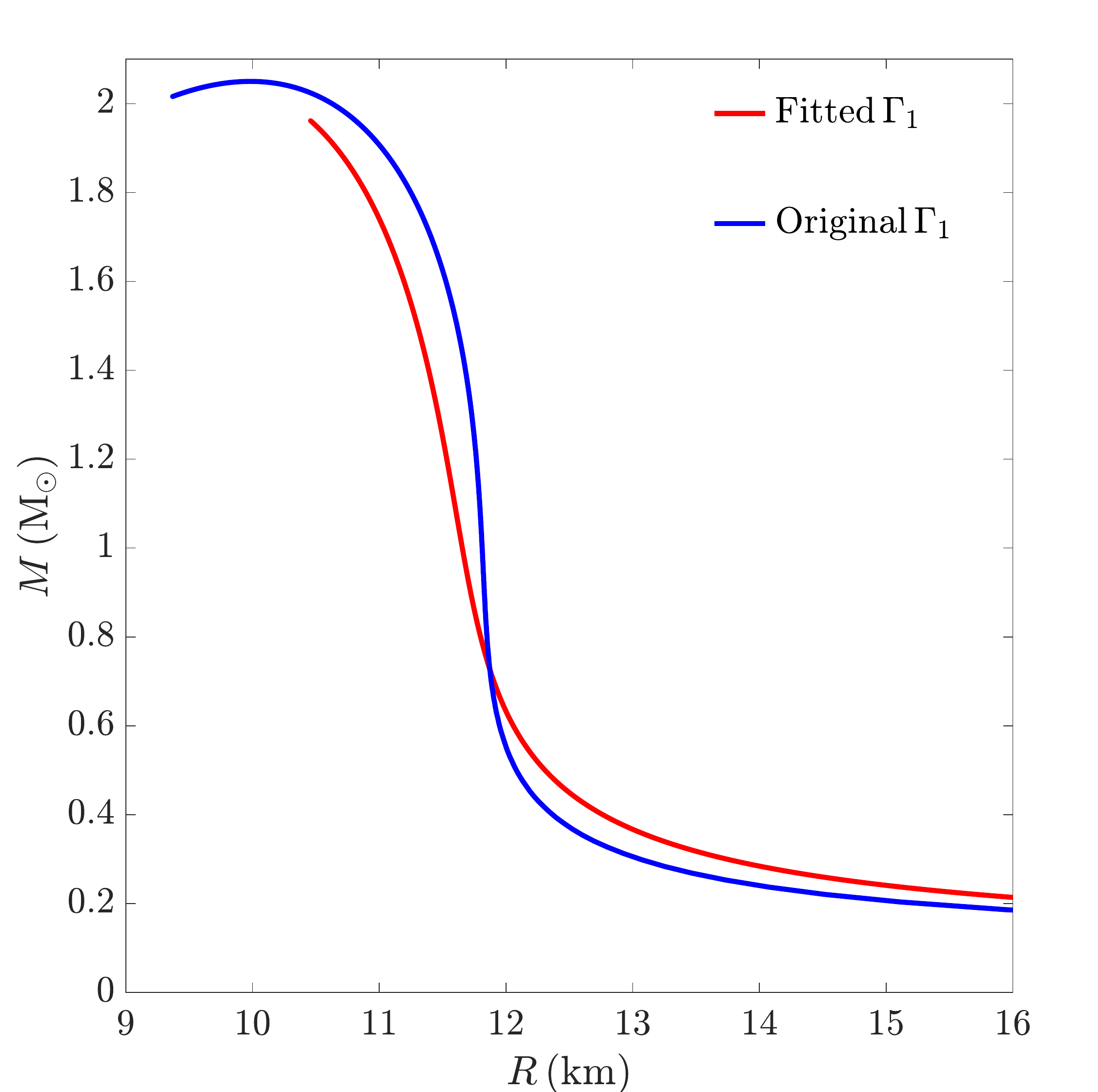}
\caption{The adiabatic index $\Gamma_1$ (upper panel) and the corresponding M-R relation (lower panel). The blue line is the $\Gamma_1$ of SLy4 EoS using a fine interpolation algorithm, while the red line is the one using a polynomial interpolation algorithm. Although these two approaches fit quite well at the low density region, they are deviated from each other in the crust-core transition region. If the accuracy of the interpolation algorithm was not well-controlled, the resultant $\Gamma_1$ actually corresponds to a different EoS hence a different TOV solution.}.
\label{fig:gamma_sensitive}
\end{figure}

$\bullet$\,\textbf{Numerical details on the adiabatic index}---
The adiabatic index $\Gamma_1$ characterizes the stiffness of the EoS, reflects some physical effects like the the softening of the EoS when the density is higher than the neutron drip density or the stiffening at the crust-core interface of SLy4 EoS. However, we need to be careful in choosing the algorithm when we obtain the adiabatic index. For example, using a 9-order polynomial interpolation to obtain the $\Gamma_1$ shown  approximately in Fig.\,\ref{fig:gamma_sensitive} will have some problems. First, this method may significantly alter the EoS stiffness behavior around the crust-core region, thereby removing some important physical facts that may affect the properties of the i-mode. Second, the altered $\Gamma_1$ essentially represents a new $P-\rho$ relation thus a different EoS. It creates an artificial inconsistency with the original background static star solution computed by the original $P-\rho$ relation. Our test shows that this inconsistency will overestimate the $i$-mode frequency from 54Hz to 170Hz due to the increasing of $\Gamma_1$ at the crust-core region, for the EoS SLy4 with 1.4 $M_{\odot}$.

\section{Crust melting}
We follow the same approach as in\,\cite{Pan2020} to compute the effect of crust melting to the binary neutron stars system, which is briefly summarised in this appendix section.

Once the tidal-induced shear strain exceeds the elastic limit $\epsilon_{b}\sim 0.1$, the plastic deformation of the crust will cause heat dissipation given by:
\begin{equation}
	\label{eq:plastic_energy}
	\dot{\epsilon}_{pl}=\frac{n_iZ^2e^2\omega_p}{a\mu \bar{N}\Gamma}e^{(-18.5\bar{\sigma}_b+\bar{\sigma}\bar{N})\Gamma},
\end{equation}
where $\mu$ is the elastic shear modulus:
\begin{equation}\label{eq:shear_modulus}
	\begin{split}
		&\mu=\frac{0.1194}{1+0.595(173/\Gamma)^2}\frac{n_i(Ze)^2}{a},\\
		&{\rm with}\quad n_i=\frac{N(1-X_n)}{A}.
\end{split}
\end{equation}
Here $N$ is the baryon number density, $X_n$ is the neutron abundance, $a=(3/4\pi n_i)^{1/3}$ is the average ion spacing, $\omega_p=({4\pi Z^2e^2 n_i}/{Am_b})^{1/2}$ is the ion plasma frequency, $\sigma=\mu \epsilon_{el}$ is the shear stress, $\bar{\sigma}=\sigma/(n_i Z^2e^2/a)$, $\Gamma=(Ze)^2/ak_BT$ is the melting parameter and for $\Gamma=\Gamma_m\approx180$ the crust undergoes a fluid-solid transition. The melting temperature for the bottom of the crust is $T\sim 1\,{\rm MeV}$. Then the energy heating rate due to the plastic deformation is\cite{Thompson_2017}:
\begin{equation}
	n_i\dot{e}_i=\sigma \dot{\epsilon}_{pl}(\sigma,T),
\end{equation}
where $e_i$ is the thermal energy per ion. The energy dissipation rate has an exponential dependence on the elastic strain\,(see Eq.\,\eqref{eq:plastic_energy}). Therefore, we can expect that at some point during the evolution, the heating will become very fast. 
For computing the heating effect $de_i=c_VdT$, we need the specific heat capacity per-ion given as\,\cite{1993ApJ...414..695C}:
\begin{equation}
\frac{c_V}{k_B}=8D_3(\alpha\eta)-6\frac{\alpha\eta}{\alpha\eta-1}+e^{\gamma\eta}\left(\frac{\gamma\eta}{e^{\gamma\eta-1}}\right)^2,
\end{equation}
where $D_3$ is the Debye integral, $\alpha=0.4$, and $\eta=\hbar\omega_p/k_BT$ with $\omega_p$ the plasma frequency and $T$ the temperature. We can get the time derivative of temperature: $\dot{T}={\mu\epsilon_{el} \dot{\epsilon}_{pl}}/{n_ic_V}$. We mainly focus on the temperature evolution at the crust base for its dominant role in the crust heat capacity.

The melting will affect the parameters of the interfacial mode, for example the damping factor $\gamma$ and the resonant frequency will depend on time, that is:
\begin{equation}
	\label{eq2}	\ddot{a}_m+\gamma(t)\dot{a}+\omega_{\alpha}^2(t)a_m=\frac{GM'W_{2m}Q}{D(t)^3}e^{-im\Phi(t)}.
\end{equation}
For example, the damping rate is:
\begin{equation}
		\gamma(t)\approx\left(2\int_{\rm crust}d^3x n_i\dot{e}_i\right)\left/\left(MR^2\sum_{m}|\dot{a}_m|^2+\omega_{\alpha}^2(t)|a_m|^2\right)\right.,
\end{equation}
where the denominator represents the mode’s total energy and the numerator represents the total energy dissipation rate in the crust.
The resonant frequency $\omega_\alpha(t)$ is assumed to be proportional to the square root of the shear modulus $\mu(t)$ in Eq.\,\eqref{eq:shear_modulus}, which also depends on the increasing temperature during the heating process. Combining the above equations, one can analyse how the melting process affects the coupling between interfacial mode and the orbital motion, on which the lower panel of Fig.\,\ref{fig:detectability} is based.

\end{document}